\documentclass[11pt]{article}
\newcommand{\version}{August 14, 2003}

\usepackage{amsmath}
\usepackage{amsgen,amstext,amsbsy,amsopn,amsthm, amssymb}

\newtheorem{thm}{Theorem}
\numberwithin{thm}{section}

\newtheorem{lem}{Lemma}
\newtheorem{prop}{Proposition}
\numberwithin{prop}{section}
\theoremstyle{definition}

\newcommand{\beq}{\begin{equation}}
\newcommand{\eeq}{\end{equation}}
\newcommand{\beqa}{\begin{eqnarray}}
\newcommand{\eeqa}{\end{eqnarray}}
\newcommand{\infspec}{{\rm inf\, spec\, }}

\newcommand{\eps}{\epsilon}

\newcommand{\V}{V}

\newcommand{\half}{\mbox{$\frac{1}{2}$}}
\newcommand{\E}{\mathcal{E}}

\newcommand{\B}{\mathcal{B}}

\newcommand{\x}{{\bf x}}
\newcommand{\y}{{\bf y}}

\newcommand{\R}{\mathbb{R}}
\newcommand{\N}{\mathbb{N}}

\newcommand{\const}{{\rm const. \,}}

\newcommand{\pmin}{\phi_{\alpha,{\rm min}}}
\newcommand{\pmax}{\phi_{\alpha,{\rm max}}}
\newcommand{\ppr}{\phi_{p,r}}

\numberwithin{equation}{section}
\begin{document}
\markboth{\scriptsize{LSY \version}}{\scriptsize{LSY \version}}
\title{\bf One-Dimensional Behavior of Dilute, Trapped Bose Gases}
\author{\vspace{5pt} Elliott H.~Lieb$^{1,*}$, Robert
Seiringer$^{1,2,\dagger}$, and Jakob Yngvason$^{2}$\\
\vspace{-4pt}\small{$1.$ Department of Physics, Jadwin Hall,
Princeton University,} \\ \small{P.~O.~Box 708, Princeton, New
Jersey
  08544}\\
\vspace{-4pt}\small{$2.$ Institut f\"ur Theoretische Physik,
Universit\"at
Wien,}\\
\small{Boltzmanngasse 5, A-1090 Vienna, Austria}}
\date{\small \version}
\maketitle

\renewcommand{\thefootnote}{$*$}
\footnotetext{Work partially
supported by U.S. National Science Foundation
grant PHY 01-39984.}
\renewcommand{\thefootnote}{$\dagger$}
\footnotetext{Erwin Schr\"odinger Fellow,
supported by the Austrian Science Fund.\\
\copyright\, 2003 by the authors. This paper may be reproduced, in its
entirety, for non-commercial purposes.}

\begin{abstract}
Recent experimental and theoretical work has shown that there are
conditions in which a trapped, low-density Bose gas behaves like the
one-dimensional delta-function Bose gas solved years ago by Lieb and
Liniger.  This is an intrinsically quantum-mechanical phenomenon
because it is not necessary to have a trap width that is the size of
an atom -- as might have been supposed -- but it suffices merely to
have a trap width such that the energy gap for motion in the
transverse direction is large compared to the energy associated with
the motion along the trap.  Up to now the theoretical arguments have
been based on variational - perturbative ideas or numerical
investigations.  In contrast, this paper gives a rigorous proof of the
one-dimensional behavior as far as the ground state energy and
particle density are concerned. There are four 
parameters involved: the
particle number, $N$, transverse and longitudinal dimensions of the
trap, $r$ and $L$, and the scattering length $a$ of the interaction
potential.  Our main result is that if $r/L\to 0$ and $N\to\infty$ the
ground state energy and density can be obtained by minimizing a
one-dimensional density functional involving the Lieb-Liniger energy
density with coupling constant $\sim a/r^2$.

This density functional simplifies in various limiting cases and we
identify five asymptotic parameter regions altogether.  Three of these,
corresponding to the weak coupling regime, 
can also be obtained as limits of a three-dimensional Gross-Pitaevskii
theory.  We also show that Bose-Einstein condensation in the ground
state persists in a part of this regime.  In the strong coupling
regime the longitudinal motion of the particles is strongly
correlated.  The Gross-Pitaevskii description is not valid in this
regime and new mathematical methods come into play.
\end{abstract}

\section{Introduction}

The technique of trapping and cooling atoms, that led to the first
realization of Bose-Einstein condensation (BEC) in dilute alkali gases
in 1995 \cite{cornell,ketterle}, has recently opened the possibility
for experimental studies, in highly elongated traps, of Bose gases
that are effectively one-dimensional.  Some of the remarkable
properties of ultracold one-dimensional Bose systems with delta
function interactions, analyzed long ago \cite{LL,L}, may thus become
accessible to experimental scrutiny in the not too distant future. 
Among these are pseudo-fermionic behavior \cite{gira}, the absence of
BEC in a dilute limit \cite{Lenard,pita,girardeau,
papenbrock,forrester}, and an excitation spectrum different from that
predicted by Bogoliubov's theory \cite{L,jackson,komineas}.  The paper
\cite{olshanii} by Olshanii triggered a number of theoretical
investigations on the transitions from 3D to an effective 1D behavior
with its peculiar properties, see, e.g., \cite{das,das2,dunjko, Ga,
girardeau2, kolomeisky, MS, petrov,tanatar}; systems showing the first
evidence of such a transition have recently been prepared
experimentally \cite{bongs,goerlitz,greiner,schreck}.

Until now the theoretical work on the dimensional cross-over in
elongated traps has either been based on variational calculations,
starting from a three-dimensional delta-potential
\cite{das2,girardeau2,olshanii}, or on numerical Monte Carlo studies
\cite{astra,blume} with more realistic, genuine 3D potentials but
particle numbers limited to the order of 100.  This work is important
and has led to valuable insights, in particular about different
parameter regions \cite{dunjko,petrov,MS},
but a more thorough theoretical understanding is clearly desirable
since this is not a simple problem.  In fact, it is evident that for a
potential with a hard core the true 3D wave function does not
approximately factorize in the longitudinal and transverse variables
and the effective one-dimensional potential can not be obtained by
simply integrating out the transverse variables of the 3D potential. 
In this sense the problem is more complicated than in a somewhat
analogous situation of atoms in extremely strong magnetic fields
\cite{BSY,LSoY}, where the Coulomb interaction behaves like an
effective one-dimensional delta potential when the magnetic field
shrinks the cyclotron radius of the electrons to zero.  In that case
the delta potential can be obtained formally by integrating out the
variables transverse to the field in a suitable scaled Coulomb
potential.  With a hard core, on the other hand, where the energy is
essentially kinetic, this method will not work since it would
immediately introduce impenetrable barriers in 1D. The one-dimensional
effective interaction emerges only if the kinetic part of the
Hamiltonian and the potential are considered {\it together}.

In the present paper we start with an {\it arbitrary}, repulsive 3D
pair potential of finite range and prove rigorously that in a well
defined limit the ground state energy and particle density of the
system are described {\it exactly} by a one-dimensional model with
delta-function interaction.  This is a highly quantum-mechanical phenomenon with no classical
counterpart, since a 1D description is possible {\it even
though the transverse trap dimension is much larger than the range of the
atomic forces.} It suffices that the energy gap associated with the
transverse confinement is much larger than the internal energy per
particle.

While the three-dimensional density remains low (in the sense that
distance between particles is large compared to the three-dimensional
scattering length) the one-dimensional density can either be high or
low.  We remark that, in contrast to three-dimensional gases, {\it
high density} in one dimension corresponds to {\it weak interactions}
and vice versa \cite{LL}.  In this paper we shall always be concerned
with large particle number, $N$, which is appropriate for the
consideration of actual experiments.  In order to make precise
statements we shall typically take the limit $N\to\infty$ but the
reader can confidently apply these limiting statement to finite
numbers like $N=100$.

Besides $N$, the parameters of the problem are the scattering length,
$a$, of the two-body interaction potential, and two lengths, $r$ and
$L$, describing the transverse and the longitudinal extension of the
trap potential, respectively.  To keep the introductory discussion
simple let us first think of the case that the particles are confined
in a box with dimensions $r$ and $L$.  The three-dimensional particle
density is then $\rho^{\rm 3D}= N/(r^2 L)$ and the one-dimensional
density $\rho^{\rm 1D}=N/L$.  The case of quadratic or more general
trapping potentials will be considered later.  We begin by
describing the division of the space of parameters into two basic
regions.  This decomposition will eventually be refined into five
regions, but for the moment let us concentrate on the basic dichotomy.

In earlier work \cite{lsy1,LSSY} we have proved that the
three-dimensional Gross-Pitaevskii formula for the energy (including
its limiting `Thomas-Fermi' case) is correct to leading order in
situations in which $a$ is small and $N$ is large.  This energy has
two parts: The energy necessary to confine the particles in the trap,
which is roughly $(\hbar^2/2m)N (r^{-2} +L^{-2})$, plus the internal
energy of interaction, which is $(\hbar^2/2m) N 4\pi a \rho^{\rm 3D}$.
The trouble is that while this formula is correct for a {\it fixed}
confining potential in the limit $N\to \infty$ with $a^3 \rho^{\rm
  3D}\to 0$, it does not hold uniformly if $r/L$ gets small as $N$
gets large.  In other words, new physics can come into play as $r/L\to
0$ and it turns out that this depends on the ratio of $a/r^2$ to
$\rho^{\rm 1D}=N/L$ .  As we shall show, the two basic regimes to
consider in highly elongated traps, i.e., when $r \ll L$, are
\begin{itemize}
\item The one-dimensional limit of the 
three-dimensional Gross-Pitaevskii/`Thomas-Fermi' regime
\item The `true' one-dimensional regime.
\end{itemize}
The former is characterized by $aL/r^2N\to 0$, while in the latter
regime $aL/r^2N$ is of the order one or even tends to infinity (which
is referred to as the Girardeau-Tonks\footnote{We call this the
Girardeau-Tonks region only because many authors refer in the present
context to Tonks \cite{tonks}. In our opinion this should really be
called the Girardeau region because it was he who first understood how
to compute the spectrum of a 1D quantum-mechanical hard core gas and
who understood that the Fermi-Dirac wave functions played a role
\cite{gira}. Tonks was interested in the positive temperature
partition functions of a hard core classical gas -- a very different
and much simpler question.} region).  These two regimes correspond to
high one-dimensional density (weak interaction) and low
one-dimensional density (strong interaction), respectively.  The
significance of the combination $aL/r^2N$ can be understood by noting
that it is the ratio of the 3D energy per particle, $\sim a\rho^{\rm
3D}\sim Na/r^2L$, to the 1D energy $\sim (\rho^{\rm 1D})^2=(N/L)^2$.
Physically, the main difference between the two regimes is that for
strong interactions the motion of the particles in the longitudinal
direction is highly correlated, while in the weak interaction regime
it is not. Mathematically, this distinction also shows up in our
proofs.
 
In both regimes the internal energy of the gas is small compared to
the energy of confinement which is of order $N/r^2$.  However,
this in itself does not imply a specifically one-dimensional behavior.
(If $a$ is sufficiently small it is satisfied in a trap of any shape.)
One-dimensional behavior, when it occurs, manifests itself by the fact
that the transverse motion of the atoms is uncorrelated while the
longitudinal motion is correlated (very roughly speaking) in the same
way as pearls on a necklace.  Thus, the true criterion for 1D behavior
is that $aL/r^2N$ is of the order unity and not merely the condition
that the energy of confinement dominates the internal energy.

The starting point for our investigations is the Hamiltonian for $N$
spinless Bosons in a confining 3D trap potential and with a short
range, repulsive pair interaction.  We find it convenient to write the
Hamiltonian in the following way (in appropriate units):
\beq\label{3dham}
H_{N,L,r,a}=\sum_{j=1}^N \left( -\Delta_j + V^{\perp}_{r}(\x^\perp_j)
+ V_{L} (z_j) \right) + \sum_{1\leq i<j\leq N} v_{a}(|\x_i-\x_j|)
\eeq 
with $\x=(x,y,z)=(\x^\perp,z)$, 
\beq
V^{\perp}_{r}(\x^\perp)=\frac 1{r^2} V^{\perp}(\x^\perp/r), \quad
V_L(z)=\frac 1{L^2} V (z/L), \quad v_{a}(|\x|)=\frac 1{a^2}v(|\x|/a) \
,
\eeq 
where $r, L, a$ are variable scaling parameters while $V^{\perp}$, $V$
and $v$ are fixed.  The interaction potential $v$ is supposed to be
nonnegative, of finite range and have scattering length 1; the scaled
potential $v_{a}$ then has scattering length $a$.  The external trap
potentials $V$ and $V^\perp$ confine the motion in the longitudinal
($z$) and the transversal ($\x^{\perp}$) directions, respectively, and
are assumed to be locally bounded and tend to $\infty$ as $|z|$ and
$|\x^{\perp}|$ tend to $\infty$.  To simplify the discussion we find
it also convenient to assume throughout that $V$ is homogeneous of
some order $s>0$, namely $V(z)=|z|^s$, but weaker assumptions,
e.g. asymptotic homogeneity \cite{lsy2}, would in fact suffice.  The
case of a simple box with hard walls is realized by taking $s=\infty$,
while the usual harmonic approximation is $s=2$. Moreover, to avoid
unnecessary technicalities we shall assume that $V^\perp$ is
polynomially bounded at infinity, but our results certainly also hold
for faster growing potentials, or even finite domains with Dirichlet
boundary conditions. Units are chosen so that $\hbar=1$ and $2m=1$.
It is understood that the lengths associated with the ground states of
$-d^2/dz^2+V(z)$ and $-\Delta^\perp+V^\perp(\x^\perp)$ are both of the
order $1$ so that $L$ and $r$ measure, respectively, the longitudinal
and the transverse extensions of the trap.  We denote the ground state
energy of (\ref{3dham}) by $E^{\rm QM}(N,L,r,a)$ and the ground state
particle density by $\rho^{\rm QM}_{N,L,r,a}(\x)$.

In parallel with the three-dimensional Hamiltonian we consider the 
Hamiltonian for $n$  Bosons in one dimension
with delta interaction
and coupling constant $g\geq 0$, i.e., 
\beq\label{13}
H_{n,g}^{\rm 1D}=\sum_{j=1}^n-\partial_j^2 + g \sum_{1\leq i<j\leq n} 
\delta(z_i-z_j)\ ,
\eeq
where $\partial_j=\partial/\partial z_{j}$. We consider this
Hamiltonian for the $z_{j}$ in an interval of length $\ell$ in the
thermodynamic limit, $\ell\to\infty$, $n\to\infty$ with $\rho=n/\ell$
fixed. The ground state energy per particle in this limit is
independent of boundary conditions and can, according to \cite{LL}, be
written as
\beq \label{1dendens}
e_{0}^{\rm 1D}(\rho)=\rho^2e(g/\rho)\ ,
\eeq
with a function $e(t)$ determined by a certain integral equation. Its
asymptotic form is $e(t)\approx
\half t$ 
for $t\ll 1$ and $e(t)\to \pi^2/3$ for $t\to \infty$. Thus
\beq\label{e0limhigh}
e_{0}^{\rm 1D}(\rho)\approx \half g\rho\ \ \hbox{\rm for}\ \ g/\rho\ll
1
\eeq
and
\beq\label{e0limlow}
e_{0}^{\rm 1D}(\rho)\approx \frac{\pi^2}3 \rho^2\ \
\hbox{\rm for}\ \
g/\rho\gg
1\ .
\eeq
Taking $\rho e_{0}^{\rm 1D}(\rho)$ as a local energy density for an 
inhomogeneous one-dimensional system we can form the energy 
functional 
\beq\label{genfunc}
\boxed{\quad \E[\rho]=\int_\R \left( |\nabla\sqrt\rho(z)|^2 +
V_{L}(z)\rho(z) + \rho(z)^3 e(g/\rho(z)) \right) dz \quad}
\eeq
with ground state energy defined by minimizing $\E[\rho]$ over all 
normalized densities $\rho$, i.e., 
\beq\label{genfuncen}
\boxed{\quad 
E(N,L,g)=\inf \left\{ \E[\rho] \, : \, \rho\geq 0\, , \, \int_\R 
\rho(z)dz = N
\right\} \ .\quad } 
\eeq
By standard methods (cf., e.g., \cite{lsy1}) one can show that there 
is a 
unique minimizer,  i.e., a density 
$\rho_{N,L,g}(z)$ with $\int \rho_{N,L,g}(z)dz=N$ and 
$\E[\rho_{N,L,g}]=
E(N,L,g)$. Here it is important to note that $t\mapsto t^3 e(1/t)$ is 
convex. 
We define the {\it mean 1D density} of this minimizer to be
\beq
\bar\rho= \frac 1N\int_\R \left(\rho_{N,L,g}(z)\right)^2 dz \ .
\eeq
In a rigid box, i.e., for $s=\infty$, $\bar \rho$ is simply $N/L$, but
in more general traps it depends also on $g$ besides $N$ and $L$.  The
order of magnitude of $\bar\rho$ in the various parameter regions will
be described in the next section.

Our main result relates the 3D ground state energy of (\ref{3dham}),
$E^{\rm QM}(N,L,r,a)$, to the 1D density functional energy $E(N,L,g)$
for a suitable $g$ in the large $N$ limit provided $r/L$ and $a/r$ are
sufficiently small. To state this precisely, let $e^\perp$ and
$b(\x^\perp)$ denote the ground state energy and the normalized,
nonnegative ground state wave function of
$-\Delta^\perp+V^\perp(\x^\perp)$, respectively. The corresponding
quantities for $-\Delta^\perp+V^\perp_{r}(\x^\perp)$ are $e^\perp/r^2$
and $b_{r}(\x^\perp)=(1/r)b(\x^\perp/r)$. In the case that the trap is
a cylinder with hard walls $b$ is a Bessel function; for a quadratic
$V^\perp$ it is a Gaussian. In any case, $b$ is a bounded function 
and, in particular, $b\in L^4(\R^2)$. Hence we can define $g$ by
\beq\label{defg}
g=\frac {8\pi a}{r^2} \int_{\R^2} |b(\x^\perp)|^4 d^2\x^\perp={8\pi a}
\int_{\R^2} |b_{r}(\x^\perp)|^4d^2\x^\perp\ .
\eeq
Our main Theorem is:

\begin{thm}[From 3D to 1D]\label{T1}
Let $N\to\infty$ and simultaneously $r/L\to 0$ and $a/r\to 0$ in such a way
that
$r^2\bar\rho\cdot\min\{\bar\rho,g\}\to 0$. Then 
\beq\label{111}
\lim \frac {E^{\rm QM}(N,L,r,a)-Ne^\perp /r^2 }{E(N,L,g)} = 1 \ .
\eeq
\end{thm}

Note that because of (\ref{e0limhigh}) and (\ref{e0limlow}) the
condition $r^2\bar\rho\cdot\min\{\bar\rho,g\}\to 0$ is the same as 
\beq \label{condition}
e_{0}^{\rm 1D}(\bar \rho)\ll 1/r^2 \ ,
\eeq
i.e., the average energy per particle associated with the longitudinal
motion should be much smaller than the energy gap between the ground
and first excited state of the confining Hamiltonian in the transverse
directions.  (The precise meaning of $\ll$ is that the ratio of the
left side to the right side tends to zero in the limit considered.) 
Note also that while the one-dimensional density can be either high or
low (compared to $g$), the gas is always {\it dilute} in 
a three-dimensional sense in the limit considered, i.e., $a^3 
\rho^{\rm 3D}\sim a^2 g \bar\rho \ll 1$.

The two regimes mentioned previously correspond to specific
restrictions on the size of the ratio $g/\bar\rho$ as $N\to\infty$,
namely $g/\bar\rho\ll 1$ for the limit of the 3D Gross-Pitaevskii
regime (weak interaction/high 1D density), and $g/\bar\rho>0$ for the
`true' one-dimensional regime (strong interaction/low 1D  density). We
shall now describe briefly the finer division of these regimes into
five regions altogether. Three of them (Regions 1--3) belong to the
weak interaction regime and two (Regions 4--5) to the strong
interaction regime. In each of these regions the general functional
(\ref{genfunc}) can be replaced by a different, simpler functional,
and the energy ${E(N,L,g)}$ in Theorem~\ref{T1} by the ground state 
energy of
that functional.

The five regions are
\begin{itemize}
\item {\bf Region 1, the Ideal Gas case:}  $g/\bar\rho\ll N^{-2}$, 
with $\bar\rho\sim N/L$,
corresponding to a non-interacting 
gas in an external potential.
\item {\bf Region 2, the 1D GP case:} 
$g/\bar\rho\sim N^{-2}$, with $\bar\rho\sim N/L$, described by a 1D 
Gross-Pitaevskii energy functional with energy density $\half 
g\rho^2$.
\item {\bf Region 3, the 1D TF case:}  $N^{-2}\ll g/\bar\rho \ll 1$, 
with 
$\bar\rho\sim (N/L) 
(NgL)^{-1/(s+1)}$, where $s$ is the degree of homogeneity of the 
longitudinal
confining potential $V$. This region is described by a Thomas-Fermi 
type functional with energy density $\half g\rho^2$, without a 
gradient 
term.
\item {\bf Region 4, the LL case:}  $g/\bar\rho\sim 1$ , with
$\bar\rho\sim  (N/L )
N^{-2/(s+2)}$, described by an energy functional with the 
Lieb-Liniger energy  (\ref{1dendens}),
without a gradient term.
\item {\bf Region 5, the GT case:} $g/\bar\rho\gg 1$, 
with $\bar\rho\sim  (N/L) N^{-2/(s+2)}$, described by a functional 
with energy density $\sim \rho^3$, corresponding to the
Girardeau-Tonks limit of the LL energy density. 
\end{itemize}
We note that the condition $g/\bar\rho\sim 1$ means that Region 4
requires the gas cloud to have aspect ratio $r/\bar L$ of the order
$N^{-1}(a/r)$ or smaller, where $\bar L\equiv N/\bar\rho \sim
LN^{2/(s+2)}$ is the length of the cloud. Experimentally, such small
aspect ratios are quite a challenge and the situations described in
\cite{bongs,goerlitz,greiner,schreck} 
are still  rather far from this regime. It may not be completely out 
of 
reach, however.

The condition $a/r\to 0$ is automatically fulfilled in Regions 1--4,
provided (\ref{condition}) holds, since
$(a/r)^2=r^2(g\bar\rho)(g/\bar\rho)=r^2\bar\rho^2(g/\bar\rho)^2\ll 1$
by (\ref{condition}) if $g/\bar\rho$ is bounded. 
 Moreover, as discusssed in the next
section, in Regions 1--2 the condition $r/L\to 0$ implies
(\ref{condition}) and hence $a/r\to 0$ , and in Region 4, $a/r\to 0$
implies (\ref{condition}).  The hypotheses of Theorem 1.1 are thus not
entirely independent.

In the next Section 2 we define the various energy functionals more
precisely and also discuss the 1D behavior of the density $\rho^{\rm
QM}_{N,L,r,a}(\x)$, separately for each region. 
 Moreover, we prove, in Subsection~\ref{3dgpsect}, that
Regions 1--3 can be reached as limiting cases of a 3D
Gross-Pitaevskii theory.  In this sense, the behavior in these regions
contains remnants of the 3D theory, which also shows up in the
fact
that BEC prevails in Regions 1 and 2, as discussed in Section 5. 
Heuristically, these traces of 3D can be understood from the fact that
in Regions 1--3 the 1D formula for the energy per particle $g\bar\rho\sim
aN/(r^2 \bar L)$, gives the same result as the three-dimensional 
formula
\cite{LY1998}, i.e., scattering length times three-dimensional
density.  This is no longer so in Regions 4 and 5 and different
mathematical methods are required.

Despite significant differences the proof of Theorem~\ref{T1} has some
basic strategies in common with the (considerably simpler) proof of
the Gross-Pitaevskii limit Theorem in \cite{lsy1} (see also 
\cite{LY1998}
and \cite{LSSY}).  The upper bound for the energy in Regions 1--3,
given in Subsection 4.1, is the simplest estimate and can be obtained by
a method analogous to that of \cite{lsy1}.  For the lower bound, and
also for the upper bound in Regions 4--5, one considers first finite
numbers of particles in boxes with Neumann or Dirichlet boundary
conditions, and subsequently puts these boxes together to treat
inhomogeneous external potentials and the infinite particle number
limit.  For the lower bound of the energy in the boxes Dyson's Lemma
\cite{dyson,LY1998,LSSY}, which converts a ``hard'' potential into a
``soft'' potential at the expense of sacrificing kinetic energy, is an
essential tool.  The main differences compared to \cite{lsy1} are on
the one hand due to the fact that in Regions 1--3 the lower bounds in
\cite{lsy1} are not valid because they are not uniform in the shape of
the trap, and on the other hand due to the correlations in the
Lieb-Liniger wave function for the longitudinal motion in Regions
4--5. For the latter reason even the proof of the upper bound in
Regions 4--5 is considerably more involved than for Regions 1--3.

Section 3 contains the main technical estimates for the boxes.  We
consider here the 3D Hamiltonian for a finite number of particles on a
finite interval in the longitudinal direction with Neumann or
Dirichlet boundary conditions and a confining potential $V^\perp_{r}$
in the transverse directions.  We estimate its energy from above and
below in terms of the energy of a 1D Hamiltonian with delta
interactions.  In Section~4 we apply these results to prove
Theorem~\ref{T1}.  In the last Section 5 we consider the question of
Bose-Einstein condensation and prove that it holds in Regions~1 and~2.

Part of the results of this work were announced in \cite{lsy3}.

\section{The five parameter regions}

We shall now discuss the simplifications of Theorem~\ref{T1} in the 
five
different parameter regions. Besides the ground state energy we also
consider the convergence of the quantum mechanical density $\rho^{\rm
QM}_{N,L,r,a}(\x)$ averaged over the transverse variables, i.e., of
\beq
\hat\rho^{\rm QM}_{N,L,r,a}(z):=\int \rho^{\rm QM}_{N,L,r,a}
(\x^\perp,z)d^2 \x^\perp \ .
\eeq
We state the results in the form of five theorems, one for each
region. These theorems will be proved in 
Section 4. Together they imply Theorem 1.1 because the energy 
functionals involved are all limiting cases of the general functional 
(\ref{genfuncen}). The
methods for proving the latter statement are fairly standard (cf., e.g.,
\cite{lsy1,lsy2} for similar computations). Since a proof of all five 
limit theorems for the functional (\ref{genfuncen}) would be 
largely repetitious, we shall limit ourselves 
to giving a  proof of two of them in Subsection~\ref{sspr}, as an 
example.

\subsection{The Ideal Gas Region}\label{igss}
This region corresponds to the trivial case where the interaction
is so weak that it effectively vanishes in the large $N$ limit and
everything collapses to the ground state of $-d^2/dz^2+V_L(z)$. By
scaling the ground state energy and wave function of this latter
operator can be written as $L^{-2}e^{\parallel }$ and
$L^{-1/2}\sqrt{\rho^{\parallel}(z/L)}$ where $e^{\parallel}$ and
$\sqrt{\rho^{\parallel}(z)}$ are the corresponding quantities for
$-d^2/dz^2+V(z)$.

\begin{thm}[Ideal gas limit]\label{T2.1}
Suppose $r/L\to 0$ and $NgL\sim NaL/r^2\to 0$ as $N\to\infty$. Then
\beq (N/L^2)^{-1}\left(E^{\rm QM}(N,L,a,r)-Ne^\perp/r^2\right)\to 
e^{\parallel}\eeq
and 
\beq
(N/L)^{-1}\hat\rho^{\rm QM}_{N,L,r,a}(Lz)\to\rho^{\parallel}(z)
\eeq
weakly in $L^1(\R)_{}$.
\end{thm}

\subsection{The 1D Gross-Pitaevskii Region}\label{gpss}
This region is described by the 1D GP density functional
\beq\label{GPfunct}
\E^{\rm GP}_{L,g}[\rho]=\int_\R \left( |\nabla\sqrt\rho(z)|^2+ 
V_L(z)\rho(z) + \half g\rho(z)^2 \right) dz
\eeq
corresponding to the high density approximation (\ref{e0limhigh}) of 
the 
interaction energy in (\ref{genfunc}). Its ground state energy
\beq\label{onedimgp}
E^{\rm GP}(N,L,g)=\inf \left\{ \E^{\rm GP}_{L,g}[\rho] \, : \, 
\rho\geq 0\, , \, \int_\R 
\rho(z)dz = N \right\}
\eeq
has the scaling property
\beq\label{1dgpscaling}
E^{\rm GP}(N,L,g)= ({N}/{L^2}) E^{\rm GP}(1,1,NgL)
\eeq
and likewise, the minimizer $\rho^{\rm GP}_{N,L,g}(z)$ satisfies
\beq
\rho^{\rm GP}_{N,L,g}(Lz)=(N/L)\rho^{\rm GP}_{1,1,NgL}(z)\ .
\eeq
\begin{thm}[1D GP limit]\label{T2.2}
Suppose $r/L\to 0$ and $NgL\sim NaL/r^2$ is fixed as 
$N\to\infty$. Then
\beq (N/L^2)^{-1}\left(E^{\rm QM}(N,L,a,r)-Ne^\perp/r ^2\right)\to 
E^{\rm GP}(1,1,NgL)\eeq
and 
\beq
(N/L)^{-1}\hat\rho^{\rm QM}_{N,L,r,a}(Lz)\to\rho^{\rm GP}_{1,1,NgL}(z)
\eeq
weakly in $L^1(\R)_{}$.
\end{thm}

\noindent{\it Remark.\/} If $r/L\to 0$ and $NgL$ stays bounded, as in 
Regions 1 and 2, condition (\ref{condition}) is automatically 
satisfied, because $g\bar\rho r^2\sim aN/L\sim (r/L)^2(NgL)$. 
Likewise, $a/r\to 0$, because $a/r=(r/L)N^{-1}(NgL)$.

\subsection{The 1D `Thomas-Fermi' Region} 

This region is a limiting case of the previous one in the sense that 
$NgL\sim NaL/r^2\to\infty$, but $a/r$ is sufficiently small so that 
$g/\bar\rho \sim 
(aL/Nr^2)(NaL/r^2)^{1/(s+1)}\to 0$, i.e.,  the high density 
approximation in (\ref{e0limhigh}) is still valid. Here $s$ is the 
degree of 
homogeneity of $V$ and the explanation of the factor 
$(NaL/r^2)^{1/(s+1)}\sim (NgL)^{1/(s+1)}$ is as follows: 
The linear extension $\bar L$ of the 
minimizing density $\rho^{\rm GP}_{N,L,g}$ is for large values of 
$NgL$ determined by $V_{L}(\bar L)\sim g(N/\bar L)$, which gives 
$\bar L\sim (NgL)^{1/(s+1)} L$.
In addition condition (\ref{condition}) requires $g\bar\rho \ll 
r^{-2}$, which 
means that $(Na/L)(NaL/r^2)^{-1/(s+1)}\to 0$.

If $NgL\sim NaL/r^2\to\infty$ the gradient term in the functional 
(\ref{onedimgp}) becomes negligible compared to the other terms. 
In fact, by a simple scaling,
\beq
\E^{\rm GP}_{L,g}[\rho]=\frac N{L^2} (NgL)^{s/(s+1)}
\int_\R \left( (NgL)^{-(s+2)/(s+1)}|\nabla\sqrt{\tilde\rho(z)}|^2+ 
V(z)\tilde\rho(z) + \half\tilde\rho(z)^2 \right) dz
\eeq
where the scaled density $\tilde \rho$ is determined by 
\beq\label{scaleddens}
\rho(z)=(N/\bar L_{\rm TF})\tilde\rho(z/\bar L_{\rm TF})\ ,
\quad \hbox{\rm with}\,\, \bar L_{\rm TF}:=(NgL)^{1/(s+1)} L \ .
\eeq
This leads to the functional
\beq\label{tffunc}
\E^{\rm TF}_{L,g}[\rho]=\int_\R \left( V_L(z)\rho(z) + \half g
\rho(z)^2 \right) dz
\eeq
whose ground state energy
\beq
E^{\rm TF}(N,L,g)=\inf \left\{ \E^{\rm TF}_{L,g}[\rho] \, : \, 
\rho\geq 0\, , \,  \int_\R 
\rho(z)dz = N \right\}
\eeq
has the scaling property
\beq\label{217}
E^{\rm TF}(N,L,g)= ({N}/{L^2}) (NgL)^{s/(s+1)}E^{\rm TF}(1,1,1)\ .
\eeq
The minimizer $\rho^{\rm TF}_{N,L,g}$ satisfies
\beq
\rho^{\rm TF}_{N,L,g}
(\bar L_{\rm TF}z)=(N/\bar L_{\rm TF})
\rho^{\rm TF}_{1,1,1}(z)
\eeq
and can be computed explicitly:
\beq\label{tfminimizer}
\rho^{\rm TF}_{1,1,1}(z)=[\mu^{\rm TF}-V(z)]_{+}
\eeq
where $[t]_{+}=\max\{t,0\}$ and $\mu^{\rm TF}$ is determined by the 
normalization $\int\rho^{\rm TF}_{1,1,1}(z)dz=1$.

Because of a formal similarity with the Thomas-Fermi energy 
functional for fermions, which also has no gradient terms, 
the functional (\ref{tffunc}) is in the literature usually 
referred to as a `Thomas-Fermi' functional, but the physics is, of 
course, quite different.

The limit theorem in this region is
\begin{thm}[1D `TF limit']\label{T2.3}
Suppose, as $N\to\infty$, $r/L\to 0$, $NgL\sim NaL/r^2\to\infty$, but
$g/\bar\rho \sim (a\bar L_{\rm TF}/Nr^2)\to 0$ and $g\bar\rho r^2\sim
Na/\bar L_{\rm TF}\to 0$ with $\bar L_{\rm TF}$ given in
(\ref{scaleddens}). Then
\beq \label{entflim}(N/L^2)^{-1}(NgL)^{-s/(s+1)}
\left[E^{\rm QM}(N,L,a,r)-Ne^\perp/r ^2\right]
\to E^{\rm TF}(1,1,1)\eeq
and 
\beq\label{dentflim}
(N/\bar L_{\rm TF})^{-1}
\hat\rho^{\rm QM}_{N,L,r,a}(\bar L_{\rm TF}z)
\to\rho^{\rm TF}_{1,1,1}(z)
\eeq
weakly in $L^1(\R)_{}$.
\end{thm}

\subsection{The Lieb-Liniger Region}
This region corresponds to the case $g/\bar\rho\sim 1$, so that
neither the high density (\ref{e0limhigh}) nor the low density
approximation (\ref{e0limlow}) is valid and the full LL energy
(\ref{1dendens}) has to be used.  The extension $\bar L$ of the system
is now determined by $V_L(\bar L)\sim (N/\bar L)^2$ which leads to
$\bar L_{\rm LL}=L N^{2/(s+2)}$, in contrast to $\bar L_{\rm TF}= L
(NgL)^{1/(s+1)}$ in the TF case.  Condition (\ref{condition}) means in
this region that $Nr/\bar L_{\rm LL}\sim N^{s/(s+2)}r/L\to 0$.  Since
$Nr/\bar L_{\rm LL}\sim(\bar\rho/g)(a/r)$, this condition is
automatically fulfilled if $g/\bar\rho$ is bounded away from zero and
$a/r\to 0$. Conversely, if $g/\bar\rho$ is 
bounded, then (\ref{condition}) implies $a/r\to 0$.

The energy functional is
\beq\label{llfunct}
\E^{\rm LL}_{L,g}[\rho]=\int_\R \left( V_{L}(z)\rho(z) + \rho(z)^3 
e(g/\rho(z)) \right) dz
\eeq
with corresponding energy
\beq
E^{\rm LL}(N,L,g)=\inf \left\{ \E^{\rm LL}_{L,g}[\rho] \, : \, 
\rho\geq 0 \, , \, \int_\R 
\rho(z)dz = N \right\}\ .
\eeq
Introducing the density parameter
\beq
\gamma:= N/\bar L_{\rm LL}=(N/L) N^{-2/(s+2)}
\eeq
we can write the scaling relation of the functional as
\beq\label{scalll}
E^{\rm LL}(N,L,g)= N\gamma^2 E^{\rm LL}(1,1,g/\gamma)\ .
\eeq
The minimizer, $\rho^{\rm LL}(z)$, satisfies
\beq\label{rhoscaling}
\rho^{\rm LL}_{N,L,g}(\bar L_{\rm LL}z)=\gamma
\rho^{\rm LL}_{1,1,g/\gamma}(z)\ .
\eeq
\begin{thm}[LL limit]\label{T2.4}
Suppose $r/L\to 0$ and $a/r\to 0$, with $g/\gamma>0$ fixed  as 
$N\to\infty$. Then
\beq 
 (N\gamma^2)^{-1}\left[E^{\rm QM}(N,L,a,r)-Ne^\perp/r ^2\right]
\to E^{\rm LL}(1,1,g/\gamma)\eeq
and 
\beq
\gamma^{-1}\hat\rho^{\rm QM}_{N,L,r,a}(\bar L_{\rm LL}z)
\to\rho^{\rm LL}_{1,1,g/\gamma}(z)
\eeq
weakly in $L^1(\R)_{}$.
\end{thm}

\subsection{The Girardeau-Tonks Region}\label{gtss}
This region corresponds to impenetrable particles, i.e, the limiting 
case $g/\bar\rho\to\infty$ and hence the formula 
(\ref{e0limlow}) for the energy density. As in Region 4, the mean 
density is here
 $\bar\rho\sim  \gamma=(N/L) N^{-2/(s+2)}$.
The energy functional is
\beq
\E^{\rm GT}_{L}[\rho]=\int_\R \left( V_{L}(z)\rho(z) + 
({\pi^2}/3 )\rho(z)^3  \right) dz
\eeq
with energy
\beq
E^{\rm GT}(N,L)=\inf \left\{ \E^{\rm GT}_{L}[\rho] \, : \, \rho\geq
0\, , \, \int_\R
\rho(z)dz = N \right\} \ ,
\eeq
which can be written
\beq
E^{\rm GT}(N,L)=  N \gamma^2 E^{\rm GT}(1,1)\ .
\eeq
The minimizer 
$\rho^{\rm GT}_{1,1}(z)$ has the form
\beq\label{gtminimizer}
\rho^{\rm GT}_{1,1}(z)=\pi^{-1}[\mu^{\rm GT}-V(z)]_{+}^{1/2} \ ,
\eeq
with $\mu^{\rm GT}$ determined by the normalization. Note that its 
shape is 
different from that of (\ref{tfminimizer}), which makes it possible 
to 
distinguish  experimentally the TF regime from the GT regime.
The scaling relation for the minimizer is
\beq
\rho^{\rm GT}_{N,L}(\bar L_{\rm LL}z)=\gamma
\rho^{\rm GT}_{1,1}(z)\ .
\eeq
\begin{thm}[GT limit]\label{T2.5}
Suppose $r/L\to 0$ and $a/r\to 0$, with $g/\gamma\to\infty$ 
as 
$N\to\infty$. Then
\beq 
 (N\gamma^2)^{-1}\left[E^{\rm QM}(N,L,a,r)-Ne^\perp/r ^2\right]
\to E^{\rm GT}(1,1)\eeq
and 
\beq
\gamma^{-1}\hat\rho^{\rm QM}_{N,L,r,a}(\bar L_{\rm LL}z)
\to\rho^{\rm GT}_{1,1}(z)
\eeq
weakly in $L^1(\R)$.
\end{thm}

\subsection{Limiting cases of the general energy functional}
\label{sspr}

As already stated, the proof of Theorem~\ref{T1} in Section 4 consists
in comparing the ground state energy of the many-body Hamiltonian
(\ref{3dham}) with the ground state energies of the functionals
defined in Subsections 2.1--2.5 in the various parameter domains. To link
these special cases to the functional (\ref{genfunc}) it then remains
to show that the ground state energy of (\ref{genfunc}) coincides with
that of the functionals in Subsections 2.1--2.5 in the appropriate
asymptotic limits. The proof of this follows the same pattern as the
derivation of the 3D TF limit from 3D GP theory in \cite{lsy1,lsy2}
and we shall here only give explicit proofs for the functionals in
Theorems 2.3 and 2.4 as examples. The limit theorems for the density
are derived from the energy convergence by variation of the external
potential in a standard way (c.f., e.g.
\cite{lsy2}).  

\begin{prop}
If $N\to\infty$, $NgL\to\infty$, but $g\bar L_{\rm TF}/N=gL(NgL)^{1/(s+1)}/N\to 0$, then
\beq\label{propen}
\left[ (N/L^2)(NgL)^{s/(s+1)} \right]^{-1}E(N,L,g)
\to E^{\rm TF}(1,1,1) 
\eeq 
and \beq\label{propdens} \left(N/\bar L_{\rm TF}
\right)^{-1}\rho_{N,L,g}\left(\bar L_{\rm TF}z\right)\to 
\rho^{\rm TF}_{1,1,1}(z) \eeq weakly in $L^1(\R)$.  
\end{prop} 
\begin{proof}
With $\tilde\rho$ the scaled density given by (\ref{scaleddens}) the 
energy
functional (\ref{genfunc}) can be written
\beqa\nonumber
\E[\rho]\!\!&=&\!\!({N}/{L^2}) (NgL)^{s/(s+1)}
 \int_\R 
\Big((NgL)^{-(s+2)/(s+1)}|\partial\sqrt{\tilde\rho(z)}|^2
\\  &&\!\! + \, 
V(z)\tilde\rho(z) +\tilde\rho(z)^3 N(g\bar L_{\rm TF})^{-1}
e(g\bar L_{\rm TF}/N \tilde\rho(z))\Big)  dz \ . 
\label{scaledfunc}
\eeqa
Now $te(1/t)\leq \half$ for all $t$ \cite{LL}, so
\beqa \nonumber 
&&\left[ (N/L^2)(NgL)^{s/(s+1)} \right]^{-1}\E[\rho]\\ \label{tfupper}
&& \leq 
\int_{\R}\left((NgL)^{-(s+2)/(s+1)}|\partial\sqrt{\tilde\rho(z)}|^2+
V(z)\tilde\rho(z) +\half\tilde\rho(z)^2\right)dz \ . 
\eeqa
Let 
\beq \label{jepsilon}
j_{\varepsilon}(z)=(2\varepsilon)^{-1}\exp(-|z|/\varepsilon) \ .
\eeq
Define $\tilde 
\rho=\rho^{\rm TF}_{1,1,1}*j_{\varepsilon}$. Then, since   $|\partial 
j_{\varepsilon}|=\varepsilon^{-1}j_{\varepsilon}$ and 
$\int\tilde\rho(z)dz=1$, 
\beq
\int|\partial\sqrt{\tilde\rho(z)}|^2dz\leq1/(4\varepsilon^2)<\infty \ 
,
\eeq
and (\ref{tfupper}) 
implies
\beq
\limsup_{N\to\infty}\left[ (N/L^2)(NgL)^{s/(s+1)} 
\right]^{-1}E(N,L,g)\leq\E^{\rm TF}_{1,1}[\rho^{\rm 
TF}_{1,1,1}*j_{\varepsilon}]
\eeq
in the limit considered. 
Moreover, $\int(\rho^{\rm 
TF}_{1,1,1}*j_{\varepsilon})^2\leq \int (\rho^{\rm 
TF}_{1,1,1})^2$ because $\int j_{\varepsilon}=1$, and 
\beq
\int\rho^{\rm 
TF}_{1,1,1}*j_{\varepsilon}(z)|z|^s dz\to \int\rho^{\rm 
TF}_{1,1,1}(z)|z|^s dz
\eeq 
for $\varepsilon\to 0$. (Note that $\rho^{\rm 
TF}_{1,1,1}=[\mu^{\rm TF}-|z|^s]_{+}$ is continuous and of compact 
support.) 
Hence
\beq\limsup_{N\to\infty}\left[ (N/L^2)(NgL)^{s/(s+1)} 
\right]^{-1}E(N,L,g)\leq E^{\rm TF}({1,1,1})\ .
\eeq
On the other hand, dropping the positive gradient term in 
(\ref{scaledfunc}) gives
\beq\label{tflower}
\left[ (N/L^2)(NgL)^{s/(s+1)} \right]^{-1}\E[\rho]\geq 
\int_\R\left(
V(z)\tilde\rho(z) +\tilde\rho(z)^3M
e(1/M\tilde\rho(z))\right) dz\ ,
\eeq
with $M=N/(g\bar L_{\rm TF})$. Note that $M\to\infty$ in the limit 
considered here. The functional on the right side of 
(\ref{tflower}) has as 
minimizer
\beq
\rho^{(M)}(z)=[f_{M}^{-1}(\mu^{(M)}-V(z))]_{+}
\eeq
where $f_{M}^{-1}$ is the inverse of the function 
$f_{M}(t)=d/dt [M t^3 e(1/tM)]$ and $\mu^{(M)}$ is chosen so that 
$\int\rho^{(M)}=1$. Note also that $t^{-1} f_M(t)\to 1$ as 
$M\to\infty$, 
uniformly on 
$[\delta,\infty[$ for every $\delta>0$. From this it follows easily 
that 
$\rho^{(M)}$ converges uniformly to $\rho^{\rm 
TF}_{1,1,1}$, given by (\ref{tfminimizer}), as $M\to\infty$. With 
$\rho=\rho_{N,L,g}$, the minimizer of $\E$, we thus obtain from 
(\ref{tflower})
\beq\label{tflower2}
\liminf_{N\to\infty}\left[ (N/L^2)(NgL)^{s/(s+1)} 
\right]^{-1}E(N,L,g)\geq 
E^{\rm TF}(1,1,1)\ .
\eeq

To 
prove the corresponding result (\ref{propdens}) for the density we 
pick a $C^\infty$ function $Y$ of compact support together with an
$\varepsilon>0$ and replace $V_{L}(z)$ 
by
\beq\label{modifiedV}
V_{L}(z)+\frac \varepsilon {L^2} 
(NgL)^{s/(s+1)}Y(L^{-1}(NgL)^{-1/(s+1)}z)=
\frac 1{L^2}(NgL)^{s/(s+1)}\left[V(z')+\varepsilon Y(z')\right]
\eeq
with $z'=z/\bar L_{\rm TF}=L^{-1}(NgL)^{-1/(s+1)}z$. While 
$V(z')+\varepsilon 
Y(z')$ is not strictly homogeneous of order $s$, it is 
asymptotically homogeneous in the sense of Def.\ 1.1 in \cite{lsy2} 
and as in the proof of Lemma 2.3 in \cite{lsy2} this is sufficient 
for 
(\ref{propen}), now with the modified external potential 
(\ref{modifiedV}).  Since both (\ref{genfuncen}) and 
the TF 
energy are concave in $\varepsilon$, the derivative with respect to 
$\varepsilon$ can be exchanged with
the limits $N\to\infty$, $(NgL)\to\infty$, giving (\ref{propdens}) in 
the sense of distributions. Since the densities have norm 1, the 
convergence holds also weakly in $L^1(\R)$.
\end{proof}

\begin{prop}
If $N\to\infty$ with $g/\gamma$ fixed, where
$\gamma=N/\bar L_{\rm LL}=(N/L)N^{-2/(s+2)}$, then
\beq
\label{propen2}(N\gamma^2)^{-1}E(N,L,g) \to E^{\rm
LL}(1,1,g/\gamma) 
\eeq
and
\beq
\gamma^{-1}\rho_{N,L,g}(\bar L_{\rm LL}z)
\to\rho^{\rm LL}_{1,1,g/\gamma}(z)
\eeq
weakly in $L^1(\R)$.
\end{prop}
\begin{proof}
With $\bar L_{\rm LL}=LN^{2/(s+2)}$ we define the scaled density
$\tilde\rho$ by 
\beq 
\rho(z)=(N/\bar L_{\rm LL})\tilde\rho(z/\bar
L_{\rm LL}) \ .  
\eeq 
The energy functional (\ref{genfunc}) can then be
written 
\beq\label{scaledfunc2}
\E[\rho]=N\gamma^2
\int_\R \left[
N^{-2}|\partial\sqrt{\tilde\rho(z)}|^2
+
V(z)\tilde\rho(z) +\tilde\rho(z)^3
e\big(g/(\gamma\tilde\rho(z))\big)\right]dz\ .
\eeq
The lower bound
\beq
(N\gamma^2)^{-1}E(N,L,g)\geq E ^{\rm LL}({1,1,g/\gamma})
\eeq
follows simply by dropping the positive gradient term from the right 
side of (\ref{scaledfunc2}).

For the upper bound take  $j_{\varepsilon}$ as in (\ref{jepsilon}) 
and define $\tilde\rho= 
j_{\varepsilon}*\rho^{\rm LL}_{1,1,g/\gamma}$ to obtain 
\beq
(N\gamma^2)^{-1}\E[\rho]\leq\frac{N^{-2}}{4\varepsilon^2}+
   \E^{\rm 
    LL}_{1,1,g/\gamma}[\tilde \rho]
\eeq
and hence
\beq
\limsup_{N\to\infty}(N\gamma^2)^{-1}E(N,L,g)\leq
   \E^{\rm 
    LL}_{1,1,g/\gamma}[j_{\varepsilon}*\rho^{\rm LL}_{1,1,g/\gamma}]
\eeq
for all $\varepsilon>0$. The convergence
\beq
\lim_{\varepsilon\to 0}\E^{\rm 
    LL}_{1,1,g/\gamma}[j_{\varepsilon}*\rho^{\rm LL}_{1,1,g/\gamma}]=
 \E^{\rm 
    LL}_{1,1,g/\gamma}[\rho^{\rm LL}_{1,1,g/\gamma}]=E ^{\rm 
    LL}({1,1,g/\gamma})  
\eeq
 follows by continuity of $|z|^s$ and $t^2e(t)$ and uniform 
convergence 
 of $j_{\varepsilon}*\rho^{\rm LL}_{1,1,g/\gamma}$ to 
 $\rho^{\rm LL}_{1,1,g/\gamma}$. 

The convergence of the densities follows as in the previous proposition
by perturbing 
the external potential, this time replacing $V_{L}(z)$ 
by
\beq\label{modifiedV2}
V_{L}(z)+\varepsilon \gamma^2 Y(L^{-1}N^{-2/(s+2)}z)=
\gamma^2\left[V(z')+\varepsilon Y(z')\right]
\eeq
with $z'=z/\bar L_{\rm LL}=L^{-1}N^{-2/(s+2)}z$. \end{proof}

\subsection{One-dimensional GP as limit of three-dimensional GP}
\label{3dgpsect}

We shall now demonstrate that the ground state energy in 
Regions 1--3 can be ob\-tained as a limit of the three-dimensional 
Gross-Pitaevskii energy. The latter is defined by the
energy functional
\beq\label{3dgp}
\E^{\rm GP}_{\rm 3D}[\Phi]=\int_{\R^3}\left(|\nabla\Phi(\x)|^2+\left\{ 
V_{r}^\perp(\x^\perp)+ V_{L}(z)\right\}|\Phi(\x)|^2+4\pi 
a|\Phi(\x)|^4\right)
d^3\x \ .
\eeq
We denote its ground state energy, i.e, the infimum over all $\Phi$ 
with 
$\int|\Phi|^2=N$,  by $E^{\rm GP}_{\rm 3D}(N,L,r,a)$. It satisfies 
the 
scaling relation
\beq\label{3dgpscaling}
E^{\rm GP}_{\rm 3D}(N,L,r,a)=(N/L^2)E^{\rm GP}_{\rm 3D}(1,1,r/L,Na/L) 
\ .
\eeq
Because of (\ref{3dgpscaling}) and (\ref{1dgpscaling}) it is 
sufficient to compare $E^{\rm GP}_{\rm 3D}$ and $E^{\rm GP}$ for 
$N=1$ and $L=1$.

\begin{thm}
Let $g$ be given by (\ref{defg}). 
In the limit $r\to 0$ and $a\to 0$, 
\beq 
\frac{E^{\rm GP}_{\rm 3D}(1,1,r,a)- e^\perp/r^2}{E^{\rm GP}(1,1,g)} 
\to 1 \ ,
\eeq
uniformly in $g$ as long as $r^2 E^{\rm GP}(1,1,g)\to 0$. 
\end{thm}

\begin{proof}
We denote the minimizer of the one-dimensional GP functional
(\ref{GPfunct}) with $N=1$, $L=1$ and $g$ fixed by $\phi(z)^2$.  
Taking
$b_{r}(\x^\perp)\phi(z)$ as trial function for the 3D functional
(\ref{3dgp}) and using the definition (\ref{defg}) of $g$ we obtain
without further ado the upper bound 
\beq 
E^{\rm GP}_{\rm
3D}(1,1,r,a)\leq e^\perp/r^2+ E^{\rm GP}(1,1,g)
\eeq 
for all $r$ and
$a$. For a lower bound we consider the one-particle Hamiltonian 
\beq
H_{r,a}=-\Delta^\perp+V_{r}^\perp(\x^\perp)-\partial_{z}^2+V(z)+8\pi
ab_{r}(\x^\perp)^2\phi(z)^2\ .  
\eeq 
Taking the 3D Gross-Pitaevskii minimizer $\Phi(\x)$ for $N=1$, $L=1$,
as trial function we get
\beqa\label{hupper} 
\inf{\,\rm spec}\, H_{r,a}&\leq& E^{\rm GP}_{\rm
3D}(1,1,r,a)-4\pi a\int \Phi^4+8\pi a\int
b_{r}^2\phi^2\Phi^2\nonumber\\ &\leq&E^{\rm GP}_{\rm 3D}(1,1,r,a)+4\pi
a\int b_{r}^4\phi^4\nonumber\\ &=&E^{\rm GP}_{\rm 3D}(1,1,r,a)+\frac
g2\int\phi^4 \ .   
\eeqa 
On the other hand, $\inf{\,\rm spec}\,H_{r,a}$ can be bounded below by
Temple's inequality \cite{TE}, which says that for any Hamiltonian $H$ with
lowest eigenvalues $E_{0}<E_{1}$ and expectation value $\langle
H\rangle<E_{1}$ in some state,
\beq\label{temple} 
E_{0}\geq\langle H\rangle-\frac{\langle(H-\langle 
H\rangle)^2\rangle}{E_{1}-\langle H\rangle}\ .
\eeq
We apply this to $H=H_{r,a}$ and the state defined by $b_{r}\phi$. 
Here 
\beq\label{expect}
\langle H\rangle=\frac{e^\perp}{r^2}+ E^{\rm GP}(1,1,g)+\frac 
g2\int\phi^4\ ,
\eeq
and since $E_{1}\geq \tilde e^\perp/r^2$ with $\tilde 
e^\perp>e^\perp$, 
(\ref{expect}) is smaller than $E_{1}$ for $r^2 E^{\rm GP}(1,1,g)$ 
small
enough. Moreover,
\beq
(H-\langle H\rangle)(b_{r}\phi)=(8\pi ab_{r}^2-g)\phi^3b_{r}
\eeq
and thus 
\beqa
\langle(H-\langle 
H\rangle)^2\rangle&=&\left(\int\phi^6\right)\left[(8\pi a)^2 
\left(\int b_{r}^6\right)-g^2\right]
\leq\left(\int\phi^6\right)(8\pi a)^2 \left(\int b_{r}^4\right)\Vert 
b_{r}\Vert_{\infty}^2\nonumber\\&
\leq &\const g\|\phi\|_\infty^2 g\int \phi^4 \leq \const E^{\rm 
GP}(1,1,g)^2 \ ,
\eeqa
where we used \cite[Lemma~2.1]{lsy2} to bound $g\|\phi\|_\infty^2$ by 
the GP energy. Combining (\ref{hupper}),
(\ref{temple}) and (\ref{expect}) we thus get
\beq
E^{\rm GP}_{\rm 3D}(1,1,r,a)-e^\perp/r^2\geq E^{\rm 
GP}(1,1,g)\left (1-\const r^2 E^{\rm GP}(1,1,g)\right)
\eeq
and the proof is complete.
\end{proof}

\noindent{\it Remark.} In combination with
Theorem 2.2 this results demonstrates {\it a fortiori} that the
three-dimensional GP limit theorem in \cite{lsy1} holds uniformly in
$r/L$, provided $NaL/r^2$ stays bounded.  A more direct proof of this
fact, closer to the lines of \cite{lsy1}, is certainly possible, but it
requires redoing all estimates keeping track of the dependence on
$r/L$.

\section{Finite  ${\mathord{\hbox{\boldmath
$n$}}}$   bounds}\label{finsect}

Before we give the proof of our main Theorem~\ref{T1} in
Section~\ref{sect4}, we will explain briefly the strategy, and give
some auxiliary results in this section. In particular, we will derive
upper and lower bounds on the ground state energy of (\ref{3dham})
with the external potential $V_L(z)$ replaced by a box with Dirichlet
and Neumann boundary conditions, respectively. To obtain bounds on the
full Hamiltonian (\ref{3dham}), space will be divided in $z$-direction
into small boxes of side length $\ell$, and the bounds of this section
will be used in every box. The reason for this is twofold: first, this
allows to consider an essentially homogeneous systems, without the
additional difficulty of the external potential $V_L(z)$, and
secondly, by varying $\ell$ the particle number in each box can be
controlled. This is necessary, since the bounds we obtain in every box
will not be uniform in the particle number.

Since the particle number in the boxes will be small (compared to
$N$), we denote it by $n$. In this section, we study
the Hamiltonian
\beq\label{defh}
H=\sum_{j=1}^n \left( -\Delta_j +
V^{\perp}_{r}(\x^\perp_j) \right) + \sum_{1\leq i<j\leq
n} v_{a}(|\x_i-\x_j|) 
\eeq 
on $L^2( (\R^2\times[0,\ell])^n )$. Let $E_{\rm D}^{\rm
QM}(n,\ell,r,a)$ and $E_{\rm N}^{\rm QM}(n,\ell,r,a)$ denote its 
ground state energy with Dirichlet and Neumann
boundary conditions, respectively. The following Theorem gives upper
and lower bounds on the ground state energy of (\ref{defh}) in terms
of the ground state energy of the 1D Hamiltonian (\ref{13}). Its proof
will be given in Subsections~\ref{upbo} and~\ref{lobo1}.  A crucial
step in the proof of the lower bound will be the use of the `Dyson
Lemma' (see \cite{dyson,LY1998}), which converts the `hard'
interaction potential $v_a$ into a `soft' one. With this new `soft' 
potential
perturbation theory can be done, and rigorous bounds are obtained with
the use of Temple's inequality (\ref{temple}).

\begin{thm}[Finite ${\mathord{\hbox{\boldmath
$n$}}}$ bounds] \label{finthm}
Let $E_{\rm D}^{\rm 1D}(n,\ell,g)$ and $E_{\rm N}^{\rm
1D}(n,\ell,g)$ denote the ground state energy of (\ref{13}) on
$L^2([0,\ell]^n)$, with Dirichlet and Neumann boundary conditions,
respectively, and let $g$ be given by (\ref{defg}). Then there is a
finite number $C>0$ such that
\beq\label{lbthm}
E_{\rm N}^{\rm QM}(n,\ell,r,a)-\frac{ne^\perp}{r^2} \geq E_{\rm 
N}^{\rm
1D}(n,\ell,g)
\left( 1 -C n 
\left(\frac{a}{r}\right)^{1/8}\left[1+\frac {nr}{\ell}
\left(\frac{a}{r}\right)^{1/8}   \right]\right)  \ .
\eeq
Moreover, 
\beq\label{ubthm}
E_{\rm D}^{\rm QM}(n,\ell,r,a)-\frac {ne^\perp}{r^2} \leq E_{\rm 
D}^{\rm
1D}(n,\ell,g)
\left(1+ C \left[ \left(\frac{n
a}{r}\right)^{2}\left( 1+ \frac {a\ell}{r^2}\right) 
\right]^{1/3}\right) \ ,
\eeq
provided the term in square brackets is less than $1$. 
\end{thm}

Let us comment briefly on the error terms in (\ref{lbthm}) and
(\ref{ubthm}). As already mentioned, in the proof of Theorem~\ref{T1}
we will divide space in the $z$-direction into small boxes of side
length $\ell$. The number of particles in each box will be roughly
$n\sim N \ell / \bar L$, where $\bar L\equiv N/\bar\rho$ the extension
of the system in $z$-direction. The $n$-dependence of the error
term in (\ref{lbthm}) restricts us essentially to have a
finite particle number $n$, i.e., that $n\sim N\ell/\bar L\sim 1$, or
$\ell\sim\bar L/N$. But for (\ref{lbthm}) to be useful we need
$\ell\gtrsim r$, i.e., $r\lesssim
\bar L/N$, or, in other words, $r$ should be of the order of the mean 
particle spacing, or smaller. For $r\gg \bar L/N$, $r$ is much bigger
than the mean particle spacing, and we have to use a different
strategy, similar to the one used in the 3D problem
\cite{LY1998,lsy1}. This will be necessary for the lower bound in
Regions 1--3. The result is stated in Theorem~\ref{finthm2}. For it's
proof it will be necessary to use the box method also in
$\x^\perp$-direction, similar to the 3D case considered in
\cite{lsy1}. However, one cannot use directly the results from there,
one has to be careful to retain uniformity in $r/L$.

Likewise, (\ref{ubthm}) will not be useful for an upper bound in all
the Regions 1--5. The reason is the last term in (\ref{ubthm}), where
we want $g\ell\sim g\bar L/N\lesssim 1$, which is only fulfilled in
Regions 1--4. For Region 5, we use a different upper bound, given in
the following Theorem. The proof of Theorem~\ref{finthm2} will be
given in Subsections~\ref{upbo2} and~\ref{lobo2}.

\begin{thm}[Additional energy bounds]\label{finthm2}
With the same notations as in Thm.~\ref{finthm},
\beqa\nonumber   
&& \!\!\!\! E^{\rm QM}_{\rm N}(n,\ell,r,a)- \frac {ne^\perp}{r^2} 
\geq\\ \nonumber
 &&\!\!\!\! \frac{ n^2 
g}{2 \ell}\left(1- C \left[n^{-1/14} + \left(\frac{n\ell 
a}{r^2}\right)^{1/8} +
\left( \left[1+ \frac{r}{\sqrt n \ell}\right] \left(\frac{\sqrt n 
a}{r}\right)^{1/4}\right)^{4/39} + \frac{na}{\ell}\right]\right)  \ . 
\\ \label{final}
\eeqa
Moreover, denoting the range of $v_a$ by $R_0$,
\beq\label{uppbou2}
E_{\rm D}^{\rm QM}(n,\ell,r,a)- \frac {ne^\perp}{r^2}\leq 
\frac{\pi^2}3
\frac{n^3}{\ell^2}\frac{\left(1+1/n\right)\left(1+1/2n
\right)}{\left(1-(n-1)R_0/\ell\right)^2} \ ,
\eeq
provided $(n-1)R_0<\ell$. 
\end{thm}

\noindent {\it Remark.} By definition, $R_0\sim a$. 
Eq. (\ref{final}) will be used with $\ell\sim (r^2 
\bar L/N)^{1/3}$ and $n\sim N\ell/\bar L$. In this case we have, in 
Regions 1--3, $n\ell g\sim a(N/r^2 \bar L)^{1/3}\ll 1$, $na/\ell\sim
a\bar\rho\ll 1$, $\sqrt n a /r \sim a(N/r^2
\bar L)^{1/3}\ll 1$ and $r/(\sqrt n \ell) \sim 1 $.

\medskip
The following four subsections contain the proofs of
Theorems~\ref{finthm} and~\ref{finthm2}. Throughout, $C$ denotes a
constant independent of the parameters, although the value of 
different
$C$'s may be different.

\subsection{Upper bound for Theorem~\ref{finthm}}\label{upbo}
In this subsection we are going to prove (\ref{ubthm}). We use the 
variational principle. 
Let $\psi$ denote the ground state of (\ref{13}) with Dirichlet 
boundary conditions, normalized by $\langle\psi|\psi\rangle=1$, and 
let $G$ and $F$ be given by
\beq
G(\x_1,\dots,\x_n)=\psi(z_1,\dots,z_n)\prod_{j=1}^n 
b_r(\x^\perp_j)
\eeq
and 
\beq
F(\x_1,\dots,\x_n)=\prod_{i<j} f(|\x_i-\x_j|) \ .
\eeq
Here $f$ is a monotone increasing function, with $0\leq f\leq 1$
and $f(t)=1$ for $t\geq R$ for some $R\geq R_0$. For $t\leq R$ we
shall choose $f(t)=f_0(t)/f_0(R)$, where $f_0$ is the solution to
the zero-energy scattering equation for $v_a$ \cite{LY1998,lsy1}. 
Note that
$f_0(R)=1-a/R$ for $R\geq R_0$, and $f'_0(t)\leq
t^{-1}\min\{1,a/t\}$. We use as a trial wave function for
(\ref{defh}) the function
\beq
\Psi(\x_1,\dots,\x_n)=G(\x_1,\dots,\x_n)F(\x_1,\dots,\x_n) \ .
\eeq
Let $\rho^{(2)}_\psi$ denote the two-particle density of $\psi$,
normalized by $\int \rho_\psi^{(2)}(z,z')dzdz'=1$. Since $F$ is $1$
whenever no pair of particles is closer together than a distance
$R$, we can estimate the norm of $\Psi$ by
\beqa\nonumber
&&\langle\Psi|\Psi\rangle \geq \int G^2
\left(1-\sum_{i<j}\theta(R-|\x_i-\x_j|)\right)\\ \nonumber &&=
1-\frac{n(n-1)}{2} \int
\rho^{(2)}_\psi(z,z')b_r(\x^\perp)^2b_r(\y^\perp)^2\theta(R-|\x-\y|)
dzdz'd^2\x^\perp d^2\y^\perp \\ \nonumber &&\geq 1-\frac{n(n-1)}{2}
\int
\rho^{(2)}_\psi(z,z')b_r(\x^\perp)^2b_r(\y^\perp)^2
\theta(R-|\x^\perp-\y^\perp|)
dzdz'd^2\x^\perp d^2\y^\perp \\ \nonumber &&= 1-\frac{n(n-1)}{2}
\int b_r(\x^\perp)^2b_r(\y^\perp)^2\theta(R-|\x^\perp-\y^\perp|)
d^2\x^\perp d^2\y^\perp \\ \label{psps} &&\geq 1- \frac{n(n-1)}2
\frac {\pi R^2}{r^2} \|b\|_4^4 \ ,
\eeqa
where we used Young's inequality \cite{anal} in the last step.

Using
\beq
\langle\Psi|-\Delta_j|\Psi\rangle=-\int F^2 G\Delta_j G + \int G^2
|\nabla_j F|^2
\eeq
and the Schr\"odinger equation $H_{n,g}\psi=E_{\rm D}^{\rm 1D}\psi$,
we can write the expectation value of (\ref{defh}) as
\beqa\nonumber
\langle\Psi|H|\Psi\rangle &=&\left(E_{\rm D}^{\rm 1D}+
\frac{n}{r^2}e^\perp\right)\langle\Psi|\Psi\rangle-g\langle\Psi|
\sum_{i<j}\delta(z_i-z_j)|\Psi\rangle
\\ \label{exp} && + \int G^2 \left( \sum_{j=1}^n |\nabla_j F|^2 +
\sum_{i<j} v_a(|\x_i-\x_j|)|F|^2\right) \ .
\eeqa
Now, since $0\leq f\leq 1$ and $f'\geq 0$ by assumption, $F^2\leq
f(|\x_i-\x_j|)^2$, and
\beq\label{fpp}
\sum_{j=1}^n |\nabla_j F|^2 \leq 2 \sum_{i<j} f'(|\x_i-\x_j|)^2 +
4 \sum_{k<i<j} f'(|\x_k-\x_i|)f'(|\x_k-\x_j|)  \ .
\eeq
Consider the first term on the right side of (\ref{fpp}), together
with the last term in (\ref{exp}). These terms are bounded above by
\beqa\nonumber
&&\!\!\!\!\!\! 2 \sum_{i<j} \int G^2 \left( f'(|\x_i-\x_j|)^2+\half
v_a(|\x_i-\x_j|) f(|\x_i-\x_j|)^2\right) \\ \nonumber
&&\!\!\!\!\!\! =n(n-1) \int
b_r(\x^\perp)^2b_r(\y^\perp)^2\rho^{(2)}_\psi(z,z')\left(
f'(|\x-\y|)^2+\half v_a(|\x-\y|) f(|\x-\y|)^2\right) . \\
\label{cha}
\eeqa
Let
\beq
h(z)=\int \left( f'(|\x|)^2+\half v_a(|\x|)
f(|\x|)^2\right)d^2\x^\perp \ .
\eeq
Note that $h(z)=0$ for $|z|\geq R$, and $\int h(z)dz= 4\pi a (1 -
a/R)^{-1}$. Using Young's
inequality for the integration over the $\perp$-variables, we get
\beq\label{31}
(\ref{cha})\leq \frac{n(n-1)}{r^2} \|b\|_4^4 \int_{\R^2}
\rho^{(2)}_\psi(z,z') h(z-z') dz dz'\ .
\eeq
Consider now the contribution from the last term in (\ref{fpp}).
We can write
\beqa\nonumber
&&\!\!\!\!\!\! 4 \sum_{k<i<j} \int G^2 
f'(|\x_k-\x_i|)f'(|\x_k-\x_j|)= \frac
23n(n-1)(n-2)  \\ \nonumber &&\!\!\!\!\!\!\cdot\int
f'(|\x_1-\x_2|)f'(|\x_2-\x_3|)
b_r(\x_1^\perp)^2b_r(\x_2^\perp)^2b_r(\x_3^\perp)^2
\rho_\psi^{(3)}(z_1,z_2,z_3)
d^3\x_1 d^3\x_2 d^3\x_3 \ , \\ \label{fppp}
\eeqa
where $\rho^{(3)}_\psi$ denotes the three-particle density of $\psi$,
normalized by 1. Let
\beq
k(z)=\int f'(|\x|)d^2\x^\perp \ ,
\eeq
which is supported in $[-R,R]$. 
For the integration over $\x_1^\perp$ we use
\beq
\int f'(|\x_1-\x_2|) b_r(\x_1^\perp)^2 d^2\x_1^\perp \leq \frac
{\|b\|_\infty^2}{r^2} k(z_1-z_2) \leq \frac {\|b\|_\infty^2}{r^2}
\|k\|_\infty \ .
\eeq
For the remaining integrations, we proceed as in (\ref{31}) to
obtain
\beq\label{ahaa}
(\ref{fppp})\leq \frac 23
n(n-1)(n-2)\frac{\|b\|_\infty^2}{r^2}\frac{\|b\|_4^4}{r^2}\|k\|_\infty
\int_{\R^2} \rho^{(2)}_\psi(z,z') k(z-z') dz
dz' \ .
\eeq
Now, for any $\phi\in H^1(\R)$,
\beqa\nonumber
\left| |\phi(z)|^2-|\phi(z')|^2\right| &=&\left| \int_{z'}^z \frac
{d|\phi(z')|^2}{dz'} dz'\right| \leq 2 \|\phi\|_\infty \int_{z'}^z
\left|\frac {d\phi(z')}{dz'}\right| dz'
\\ \label{39} &\leq& 2 |z-z'|^{1/2} \left(\int_\R
|\phi|^2\right)^{1/4}\left(\int_\R \left|\frac
{d\phi}{dz}\right|^2\right)^{3/4} \ ,
\eeqa
where we used $\|\phi\|_\infty^2\leq \|d\phi/dz\|_2 \|\phi\|_2$.
Applying this to $\rho_\psi^{(2)}(z,z')$, considered as a function of 
$z$ only, and using the fact that the support of $h$ is contained in $[-R,R]$, 
we therefore get
\beqa\nonumber
&&\!\!\!\!\!\!\int_{\R^2} \rho^{(2)}_\psi(z,z') h(z-z') dz
dz'- \int_\R h(z) dz \int \rho_\psi^{(2)}(z,z)dz \\ \nonumber 
&&\!\!\!\!\!\! \leq 2
R^{1/2} \int_\R h(z) dz \int \left |\partial_z 
\sqrt{\rho_\psi^{(2)}(z,z')}\right|^2  dz dz' \leq 2
R^{1/2} \int_\R h(z) dz 
\left\langle\psi\left|-\frac{d^2}{dz_1^2}\right|\psi\right
\rangle^{3/4} \ \!\!, \\ \label{cha2}
\eeqa
where we used Schwarz' inequality, the normalization of
$\rho^{(2)}_\psi$ and the symmetry of $\psi$. The same argument is
used for (\ref{ahaa}) with $h$ replaced by $k$. Now $\int h(z)dz=
4\pi a (1 - a/R)^{-1}$, and
\beqa\nonumber
\|k\|_\infty &\leq& \frac {2\pi a}{1-a/R} \left(1+\ln(R/a)\right)
\ ,
\\ \int_\R k(z) dz &\leq& \frac{ 2\pi aR}{1-a/R}\left(1-\frac
a{2R}\right) \ .
\eeqa
Therefore
\beq\label{bb1}
(\ref{31})+(\ref{ahaa}) \leq \frac {n(n-1)}{2} g \frac
{1+K}{1-a/R}\left[ \int \rho_\psi^{(2)}(z,z)dz + 2
R^{1/2}\left\langle\psi\left|-\frac{d^2}{dz_1^2}\right|\psi
\right\rangle^{3/4}\right]
\ ,
\eeq
where we denoted
\beq
K=\frac{2\pi}3 (n-2) \frac aR \frac {1+\ln(R/a)}{1-a/R}\left(\frac
Rr\right)^2 \|b\|_\infty^2 \ .
\eeq

It remains to bound the second term in (\ref{exp}). We use again the fact
that $F$ is equal to 1 as long as the particles are not within a
distance $R$. We obtain
\beqa\nonumber
&&\langle\Psi|\sum_{i<j}\delta(z_i-z_j)|\Psi\rangle\\ \nonumber
&&\geq \frac{n(n-1)}{2} \int \rho_\psi^{(2)}(z,z) dz  \int
b_r(\x^\perp)^2b_r(\y^\perp)^2\left(1-\theta(R-|\x^\perp-\y^\perp|)\right)
\\ &&\geq \frac{n(n-1)}{2} \int \rho_\psi^{(2)}(z,z) dz 
\left(1-\frac{\pi
R^2}{r^2} \|b\|_4^4\right) \label{cha3} \ .
\eeqa

Putting together the bounds (\ref{psps}), (\ref{bb1}) and
(\ref{cha3}), and using the fact that $g\half n(n-1)\int
\rho^{(2)}_\psi(z,z)dz\leq E_{\rm D}^{\rm 1D}$ and
$\langle\psi|-d^2/dz_1^2|\psi\rangle\leq E_{\rm D}^{\rm 1D}/n$, we
obtain the upper bound
\beqa\nonumber
&&\!\!\!\!\!\frac{\langle\Psi|H|\Psi\rangle}{\langle\Psi|\Psi\rangle}-\frac
{ne^\perp}{r^2}\leq E_{\rm D}^{\rm 1D}(n,\ell,g)\left[1-
\frac{n(n-1)}2  \frac {\pi R^2}{r^2} \|b\|_4^4\right]^{-1}\\ 
&&\!\!\!\!\!\cdot\left 
(1+\frac
{a/R+K}{(1-a/R)}+  \frac{\pi
R^2}{r^2}\|b\|_4^4+g(n-1)R^{1/2}\left(\frac n{E^{\rm
1D}_{\rm D}}\right)^{1/4}\frac {1+K}{1-a/R}\right) \ \!\! ,
\eeqa
provided the term in square brackets is positive. 
We now use $E_{\rm 
D}^{\rm
1D}/n\geq (\pi/\ell)^2$, and choose
\beq\label{Rin3.1}
R^3 = \frac {a r^2} {n^2(1+ g \ell )} \ .
\eeq
This gives (under the assumption $(na/r)^2(1+g\ell)\leq 1$)
\beq\label{uppbou}
E_{\rm D}^{\rm QM}(n,\ell,r,a)-\frac{ne^\perp}{r^2}\leq
E_{\rm D}^{\rm 1D}(n,\ell,g) \left(1+ C \left(\frac{n
a}{r}\right)^{2/3}(1+ g\ell)^{1/3}\right) 
\eeq
for some constant $C>0$.

\subsection{Lower bound for Theorem~\ref{finthm}}\label{lobo1}

We are now going to prove (\ref{lbthm}). 
We write a general wave function $\Psi$ as
\beq
\Psi(\x_1,\dots,\x_n)=f(\x_1,\dots,\x_n)\prod_{k=1}^n
b_r(\x^\perp_k) \ ,
\eeq
which can always be done, since $b_r$ is a strictly positive 
function. 
Partial integration gives
\beq\label{329}
\langle\Psi|H|\Psi\rangle=\frac{n e^\perp}{r^2}+
\sum_{i=1}^n \int \left[ |\nabla_i f|^2 +\half \sum_{j,\, j\neq i}
v_a(|\x_i-\x_j|)|f|^2 \right] \prod_{k=1}^n b_r(\x^\perp_k)^2 d^3
\x_k \ . 
\eeq
Choose some $R>R_0$, fix $i$ and $\x_j$, $j\neq i$, and consider
the Voronoi cell $\Omega_j$ around particle $j$, 
i.e.,
\beq \Omega_j=\{\x:\, |\x-\x_{j}|\leq  |\x-\x_{k}| \hbox{ for all } 
k\neq j\}. 
\eeq
Denote by $\B_j$
the ball of radius $R$ around $\x_j$. We can estimate
\beqa\nonumber
&&\int_{\Omega_j\cap \B_j} b_r(\x^\perp_i)^2\left( |\nabla_i f|^2+
\half v_a(|\x_i-\x_j|)|f|^2\right) d^3\x_i\\ \nonumber && \geq
\min_{\x\in \B_j}b_r(\x^\perp)^2 a \int_{\Omega_j\cap \B_j}
U(|\x_i-\x_j|)|f|^2 \\ && \geq \frac{\min_{\x\in
\B_j}b_r(\x^\perp)^2}{\max_{\x\in \B_j}b_r(\x^\perp)^2} a
\int_{\Omega_j\cap \B_j} b_r(\x^\perp_i)^2 U(|\x_i-\x_j|)|f|^2 \ ,
\eeqa
where we used Lemma 1 of \cite{LY1998}, and
\beq\label{defu}
U(r)=\left\{\begin{array}{ll } 3(R^3-R_0^3)^{-1} & {\rm for\
}R_0\leq r\leq R
\\ 0 & {\rm otherwise} \ .
\end{array} \right.
\eeq
For some $\delta>0$ define $\B_\delta\subset\R^2$ by
\beq
\B_\delta=\left\{ \x^\perp\in \R^2 \, :\, b(\x^\perp)^2\geq
\delta\right\} \ .
\eeq
Estimating $\max_{\x\in \B_j}b_r(\x^\perp)^2\leq
\min_{\x\in \B_j}b_r(\x^\perp)^2 + 2(R/r^3) \|\nabla
b^2\|_\infty$, we obtain
\beq\label{334}
\frac{\min_{\x\in \B_j}b_r(\x^\perp)^2}{\max_{\x\in
\B_j}b_r(\x^\perp)^2}\geq
\chi_{\B_\delta}(\x^\perp_j/r)\left(1-2\frac R{r} \frac{\|\nabla
b^2\|_\infty}{\delta}\right) \ .
\eeq
(For a proof that $\nabla b^2$ is a bounded function, see the proof of
Lemma~\ref{L5} in the Appendix). Here $\chi_{\B_\delta}$ denotes the
characteristic function of $\B_\delta$. Denoting $k(i)$ the nearest
neighbor to particle $i$, we conclude that, for $0\leq \eps\leq 1$,
\beqa\nonumber
&&\sum_{i=1}^n \int \left[ |\nabla_i f|^2 +\half \sum_{j,\, j\neq
i} v_a(|\x_i-\x_j|)|f|^2 \right] \prod_{k=1}^n b_r(\x^\perp_k)^2
d^3 \x_k \\ \nonumber && \geq \sum_{i=1}^n \int \Big[ \eps
|\nabla_i f|^2 +(1-\eps)|\nabla_i f|^2 \chi_{\min_k |z_i-z_k|\geq
R}(z_i)\\ \label{negl} &&  \quad\qquad\qquad + a'
U(|\x_i-\x_{k(i)}|)\chi_{\B_\delta}(\x^\perp_{k(i)}/r)|f|^2 \Big]
\prod_{k=1}^n b_r(\x^\perp_k)^2 d^3 \x_k \ ,
\eeqa
where $a'=a(1-\eps)(1-2 R \|\nabla b^2\|_\infty/ r \delta)$.

Define $F(z_1,\dots,z_n)\geq 0$ by
\beq\label{defF}
|F(z_1,\dots,z_n)|^2=\int |f(\x_1,\dots,\x_n)|^2\prod_{k=1}^n
b_r(\x_k^\perp)^2 d^2\x_k^\perp \ .
\eeq
Neglecting the kinetic energy in $\perp$-direction in the second term
in (\ref{negl}) and using the Schwarz inequality to bound the
longitudinal kinetic energy of $f$ by the one of $F$, we get the
estimate
\beqa\nonumber
&&\langle\Psi|H|\Psi\rangle- \frac{n e^\perp}{r^2}\geq \\ \nonumber
&& \sum_{i=1}^n \int\Big[\eps |\partial_i F|^2 +
(1-\eps)|\partial_i F|^2 \chi_{\min_k |z_i-z_k|\geq R}(z_i)
\Big]\prod_{k=1}^n dz_k \\ \nonumber &&+\sum_{i=1}^n \int \left[
\eps|\nabla^\perp_i f|^2 + a'
U(|\x_i-\x_{k(i)}|)\chi_{\B_\delta}(\x^\perp_{k(i)}/r)|f|^2
\right] \prod_{k=1}^n b_r(\x^\perp_k)^2 d^3 \x_k \ , \\ \label{57}
\eeqa
where $\partial_j=d/dz_j$, and $\nabla^\perp$ denotes the gradient
in $\perp$-direction. We now investigate the last term in
(\ref{57}). Consider, for fixed $z_1,\dots,z_n$, the expression
\beq\label{44}
\sum_{i=1}^n \int \left[ \eps|\nabla^\perp_i f|^2 + a'
U(|\x_i-\x_{k(i)}|)\chi_{\B_\delta}(\x^\perp_{k(i)}/r)|f|^2
\right] \prod_{k=1}^n b_r(\x^\perp_k)^2 d^2 \x_k^\perp \ .
\eeq
To estimate this term from below, we use Temple's inequality, as in
\cite{LY1998}. Let $\widetilde e^\perp$ denote the gap above zero
in the spectrum of $-\Delta^\perp+V^\perp-e^\perp$, i.e., the lowest
non-zero eigenvalue. By scaling, $\widetilde e^\perp/r^2$ is the gap
in the spectrum of $-\Delta^\perp+V^\perp_r-e^\perp/r^2$. Note that
under the transformation $\phi \mapsto b_r^{-1} \phi$ this latter
operator is unitarily equivalent to $\nabla^{\perp *}\cdot
\nabla^\perp$ as an operator on 
$L^2(\R^2,b_r(\x^\perp)^2 d^2\x^\perp)$, as considered in
(\ref{44}). Hence also this operator has $\widetilde e^\perp/r^2$ as
its energy gap. Denote
\beq
\langle U^k\rangle=  \int \left(\sum_{i=1}^n
U(|\x_i-\x_{k(i)}|)\chi_{\B_\delta}(\x^\perp_{k(i)}/r)\right)^k
\prod_{k=1}^n b_r(\x^\perp_k)^2 d^2 \x_k^\perp \ .
\eeq
Temple's inequality (\ref{temple}) implies (under the assumption that 
the
denominator in the last term is positive)
\beq\label{60}
(\ref{44})\geq |F|^2 a'\langle U \rangle \left(1- a'\frac{\langle
U^2\rangle}{\langle U\rangle}\frac{1}{\eps \widetilde
e^\perp/r^2-a'\langle U\rangle}\right) \ .
\eeq
Now, using (\ref{defu}) and Schwarz' inequality, $\langle
U^2\rangle\leq 3n(R^3-R_0^3)^{-1}\langle U\rangle$, and
\beqa\nonumber
\langle U\rangle&\leq& n(n-1)\int
U(|\x-\y|)b_r(\x^\perp)^2b_r(\y^\perp)^2 d^2\x^\perp d^2\y^\perp
\\ &\leq& n(n-1)\frac{\|b\|_4^4}{r^2} \int U(|\x|) d^2\x^\perp\leq
n(n-1)\frac{\|b\|_4^4}{r^2}\frac{3\pi R^2}{R^3-R_0^3} \ .
\eeqa
Using this and $a'\leq a$ in the error term, we obtain
\beq
(\ref{60})\geq |F|^2 a''\langle U \rangle \ ,
\eeq
where
\beq
a'' = a' \left(1-\frac {3n}{\eps\widetilde e^\perp}
\frac{ar^2}{R^3} \frac 1{1-(R_0/R)^3}
\left[1-\frac{n^2}{\eps\widetilde e^\perp}
\frac{a}{R}3\pi\|b\|_4^4 \frac 1{1-(R_0/R)^3}\right]^{-1}\right) \
,
\eeq
with the understanding that the term in square brackets is positive. 
Now let
\beq
d(z-z')=\int_{\R^4} b_r(\x^\perp)^2 b_r(\y^\perp)^2 U(|\x-\y|)
\chi_{\B_\delta}(\y^\perp/r) d^2\x^\perp d^2\y^\perp \ .
\eeq
Note that $d(z)=0$ if $|z|\geq R$.
We estimate $\langle U\rangle$ from below by
\beqa\nonumber
\langle U\rangle\!\! &\geq&\!\! \sum_{i\neq j} \int 
U(|\x_i-\x_j|)\chi_{\B_\delta}(\x_j^\perp/r) \prod_{k, \, k\neq i,j} 
\theta(|\x_k-\x_i|-R) \prod_{l=1}^n b_r(\x^\perp_l)^2 d^2\x^\perp_l  
\\ \nonumber 
&\geq& \!\! \sum_{i\neq j} \int 
U(|\x_i-\x_j|)\chi_{\B_\delta}(\x_j^\perp/r)\left(1-\!\!\! \sum_{k, 
\, k\neq i,j} \theta(R-|\x_k-\x_i|)\right)\prod_{l=1}^n 
b_r(\x^\perp_l)^2 d^2\x^\perp_l \\ &\geq& \!\!
\sum_{i\neq j}d(z_i-z_j) \left(1-(n-2)
\frac{\pi R^2}{r^2} \|b\|_\infty^2\right) \ .
\eeqa
Let
\beq\label{appp}
a'''=a'' \left(1-(n-2) \frac{\pi R^2}{r^2} \|b\|_\infty^2\right) \ ,
\eeq
and denote $g'=2 a''' \int_\R d(z) dz$. Note that since
$|b(\x^\perp)^2-b(\y^\perp)^2|\leq R \|\nabla b^2\|_\infty$ for
$|\x^\perp-\y^\perp|\leq R$,
\beqa\nonumber
\int_\R d(z) dz &\geq& \frac {4\pi}{r^2} \left(\int_{\B_\delta}
b(\x^\perp)^4 d^2\x^\perp - R\|\nabla b^2\|_\infty/r\right)\\
&\geq& \frac {4\pi}{r^2}\left(\|b\|_4^4- \delta- R\|\nabla
b^2\|_\infty/r\right) \ .
\eeqa
We write
\beqa\nonumber
&& \sum_{i=1}^n \int  \left[ \eps |\partial_i F|^2 + a'''
\sum_{j,\, j\neq i} d(z_i-z_j) |F|^2\right] \prod_{k=1}^n dz_k\\
\label{347} &&= \sum_{i\neq j} \int \left[ \frac\eps{n-1}
|\partial_i F|^2 + a'''  d(z_i-z_j) |F|^2\right]  \prod_{k=1}^n
dz_k\ .
\eeqa
Now consider, for fixed $z_j$, $j\neq i$, the expression
\beq\label{claim}
\int \left[ \frac\eps{n-1} |\partial_i F|^2 + a'''  d(z_i-z_j)
|F|^2\right] dz_i \ .
\eeq
We claim that
\beq\label{claim2}
(\ref{claim})\geq \half g' \max_{|z_i-z_j|\leq R}
|F|^2 \chi_{[R,\ell-R]}(z_j) 
\left(1-\left(\frac{2(n-1)}{\eps}g'R\right)^{1/2}\right) \ .
\eeq
Assume that (\ref{claim2}) is false. Estimating, for any
$H^1([0,\ell])$-function $\phi$,
\beq
|\phi(z)|^2- \max_{|z-z_0|\leq R}|\phi(z)|^2\geq - \left|
\int_z^{z_0} \partial |\phi|^2\right| \geq - 2 R^{1/2}
\max_{|z-z_0|\leq R}|\phi(z)| \left(\int_0^\ell |\partial
\phi|^2\right)^{1/2}
\eeq
and applying this estimate to $F$
(considered only as a function of $z_i$), we obtain, using $\eps
\int |\partial_i F|^2 \leq \half (n-1) g' \max_{|z_i-z_j|\leq R}
|F|^2$ by assumption,
\beqa\nonumber
&&a'''\int d(z_i-z_j) |F|^2 dz_i \\ \nonumber
&&\geq
a'''\int d(z-z_j) dz \max_{|z_i-z_j|\leq
R}|F|^2 \left(1-\left(\frac 2\eps(n-1)g'R\right)^{1/2}\right)
\\  &&\geq
 \half g'\chi_{[R,\ell-R]}(z_j) \max_{|z_i-z_j|\leq
R}|F|^2 \left(1-\left(\frac 2\eps(n-1)g'R\right)^{1/2}\right) \ , 
\label{351}
\eeqa
contradicting our assumption. This proves (\ref{claim2}).

Putting everything together, we thus obtain
\beqa\nonumber
\langle\Psi|H|\Psi\rangle- \frac{ne^\perp}{r^2} &\geq& \sum_{i=1}^n 
\int\Big[(1-\eps)|\partial_i F|^2 \chi_{\min_k
|z_i-z_k|\geq R}(z_i) \Big] \prod_{k=1}^n dz_k \\
\label{putt} & & +\sum_{i\neq j} \half g'' \int
\max_{|z_i-z_j|\leq R}|F|^2 \chi_{[R,\ell-R]}(z_j) \prod_{k,\, k\neq 
i}dz_k \ ,
\eeqa
where
\beq\label{gpp}
g''=g'\left(1-\left(\frac 2\eps(n-1)g'R\right)^{1/2}\right) \ .
\eeq
Assume that $(n+1)R<\ell$. Given an $F$ with $\int |F|^2 dz_1\cdots 
dz_n=1$, define, for
$0\leq z_1\leq z_2\leq \dots \leq z_n\leq \ell-(n+1)R$,
\beq
\psi(z_1,\dots,z_n)=F(z_1+R,z_2+2 R,z_3+3R,\dots,z_n+n R) \ ,
\eeq
and extend the function to all of $[0,\ell-(n+1)R]^n$ by symmetry. A
simple calculation shows that, for
\beq
H'=(1-\eps) \sum_{i=1}^n -\partial_i^2 + g''
\sum_{i<j}\delta(z_i-z_j)
\eeq
on $L^2([0,\ell-(n+1)R]^n)$,
\beqa\nonumber
(\ref{putt})\geq \langle\psi |H' |\psi\rangle &\geq& (1-\eps)
E_{\rm N}^{\rm 1D}(n,\ell-(n+1)R,g'') \langle\psi|\psi\rangle  \\ 
\label{puutt}
&\geq&
(1-\eps) E_{\rm N}^{\rm 1D}(n,\ell,g'') \langle\psi|\psi\rangle\ .
\eeqa
Here $\langle\psi |H' |\psi\rangle$ is interpreted in the quadratic
from sense, since $\psi$ does not necessarily fulfill Neumann boundary
conditions. Since these give the lowest energy for the quadratic form,
(\ref{puutt}) is valid anyway.

It remains to estimate $\langle\psi|\psi\rangle$ for the $F$ that
is related to the true ground state $\Psi$ by (\ref{defF}). 
We have 
\beqa\nonumber
\langle\psi|\psi\rangle\!\!&=&\!\! \int |F|^2 \prod_{k=1}^n 
\chi_{[R,\ell-R]}(z_k)\prod_{i<j}\theta(|z_i-z_j|-R) \\ \nonumber 
&\geq&\!\! 1- \int |F|^2\left[ \sum_{k=1}^n 
\left(1-\chi_{[R,\ell-R]}(z_k)\right) + \sum_{i<j} 
\theta(R-|z_i-z_j|)\chi_{[R,\ell-R]}(z_j) \right]  \ \! . \\ 
\label{rs}
\eeqa
To bound the second term on the right side of (\ref{rs}), we use
\beqa\nonumber
\sum_{i<j} \int |F|^2 
\theta(R-|z_i-z_j|)\chi_{[R,\ell-R]}(z_j) \!\! &\leq&\!\!
  R \sum_{i\neq j}\int \max_{|z_i-z_j|\leq R} |F|^2
  \chi_{[R,\ell-R]}(z_j) \\ &\leq&\!\! \frac {2R}{g''} \left(E^{\rm
  QM}_{\rm N}(N,\ell,r,a)-\frac {ne^\perp}{r^2}\right) \ \! ,
\eeqa
where the last inequality follows from (\ref{putt}). 
For the other term, we use the simple fact that, for any function 
$\phi\in H^1([0,\ell])$, and for $0\leq R\leq \ell$, 
\beq
\int_0^R \phi(z)dz - \frac R\ell \int_0^\ell \phi(z)dz = \int_0^L 
\phi'(z) f_{R,\ell}(z)dz \ ,
\eeq
with $f_{R,\ell}(z)=zR/\ell-\min\{z,R\}$. Note that 
$|f_{R,\ell}(z)|\leq R$. Applying this to $F^2$ and using Schwarz' 
inequality we get the estimate
\beqa\nonumber
\int |F|^2\sum_{k=1}^n \left(1-\chi_{[R,\ell-R]}(z_k)\right) &\leq& 2 
n \frac R\ell + 4 n R \left(\frac 1n \langle  F | \mbox{$ \sum_i$} 
-\partial_i^2 | F\rangle\right)^{1/2} \\ \nonumber &\leq& 2 n \frac 
R\ell + 4 n R \left(\frac 1n E^{\rm QM}_{\rm N}(n,\ell,r,a)-\frac 
{e^\perp}{r^2} \right)^{1/2}   \ .
\\ 
\eeqa

Denoting $A\equiv E^{\rm QM}_{\rm N}(n,\ell,r,a)-ne^\perp/r^2$, we 
thus have
\beq
A\geq E_{\rm N}^{\rm 1D}(n,\ell,g'')\left(1-\frac {2R}{g''} A - 2n 
\frac R\ell - 4 \sqrt n R A^{1/2}\right) \ ,
\eeq
which implies
\beq
A\geq \frac{E_{\rm N}^{\rm 1D}(n,\ell,g'')}{1+ {2R}E_{\rm N}^{\rm 
1D}(n,\ell,g'')/g''} \left(1- 2n\frac R\ell - 2 R \sqrt{n E_{\rm 
N}^{\rm 1D}(n,\ell,g'')}\right) \ .
\eeq
We now use the simple upper bounds
\beq
E_{\rm N}^{\rm 1D}(n,\ell,g'')\leq \frac{n(n-1)}{2\ell}g'' \quad {\rm 
and \quad } 
E_{\rm N}^{\rm 1D}(n,\ell,g'')\leq \frac{\pi^2}{3} \frac{n^3}{\ell^2} 
\ ,
\eeq
which follow from a constant and a free-fermion trial wave function, 
respectively. Moreover, by concavity of $E_{\rm N}^{\rm 1D}$ in $g$,  $E_{\rm N}^{\rm 1D}(n,\ell,g'')\geq E_{\rm 
N}^{\rm 1D}(n,\ell,g)g''/g$ for $g''\leq g$, and therefore
\beq
E^{\rm QM}_{\rm N}(n,\ell,r,a)-\frac{ne^\perp}{r^2} \geq E_{\rm 
N}^{\rm 1D}(n,\ell,g)\frac { g''}g \left(1-\frac R\ell \left[ n(n+1) 
+ \frac {2\pi}{\sqrt 3} n^2\right]\right) .
\eeq
We now choose
\beq\label{chopara}
R=r\left(\frac{a}{r}\right)^{1/4}\ , \quad
\eps=\left(\frac{a}{r}\right)^{1/8}\ , \quad
\delta=\left(\frac{a}{r}\right)^{1/8} \ ,
\eeq
and obtain
\beq\label{lowbou}
E_{\rm N}^{\rm QM}(n,\ell,r,a)-\frac{ne^\perp}{r^2} \geq 
E_{\rm N}^{\rm 1D}(n,\ell,g)\left(1-C n 
\left(\frac{a}{r}\right)^{1/8}\left[1+ \frac{n 
r}{\ell}\left(\frac{a}{r}\right)^{1/8} 
\right]\right) 
\eeq
for some constant $C>0$. 

\subsection{Upper bound for the Girardeau-Tonks regime}\label{upbo2}
We use as a trial function
\beq
\Psi(\x_1,\dots,\x_n)=\psi(z_1,\dots,z_n)\prod_{k=1}^n
b_r(\x^\perp_k) \ ,
\eeq
where $\psi$ is the ground state of a one-dimensional Bose gas of
particles with hard cores of radius $R_0$. It is well known that it's
energy is the same as that of $n$ non-interacting fermions on a
line of reduced length $\ell-(n-1)R_0$. An explicit calculation yields
(\ref{uppbou2}).

\subsection{Lower bound for Theorem~\ref{finthm2}}\label{lobo2}

We start by defining an auxiliary 2D GP energy functional by
\beq\label{aux}
\phi \mapsto \int_{\R^2} \left( |\nabla^\perp \phi|^2 + V^\perp 
|\phi|^2 + p |\phi|^4\right) d^2\x^\perp 
\eeq
for some parameter $p\geq 0$. The following fact will be needed below.

\begin{lem}\label{L5} 
For any $p\geq 0$, there exists a unique minimizer (up to a constant
phase factor) of (\ref{aux}) under the normalization condition $\int
|\phi|^2 = 1$, denoted by $\phi_p$, that can be chosen strictly
positive. Moreover, both $\phi_p$ and $\nabla^\perp \phi_p$ are
bounded uniformly in $p$ and $\x^\perp$ for $p$ in a finite interval 
$[0,P]$.
\end{lem}

The proof of this Lemma, as well as the proof of Lemmas~\ref{L2} 
and~\ref{L4} below, will be given in the Appendix. 

The energy corresponding to the minimizer $\phi_p$ will be denoted by
$E^{\rm aux}(p)$. Define also $\phi_{p,r}(\x^\perp)=r^{-1}
\phi_p(\x^\perp/r)$. Writing the wave function as
\beq
\Psi(\x_1,\dots,\x_n)=f(\x_1,\dots,\x_n)\prod_{k=1}^n
\ppr(\x^\perp_k) \ ,
\eeq
using partial integration and the variational equation for $\phi_p$, 
we obtain
\beqa\nonumber
&&\!\!\!\!\!\langle\Psi|H|\Psi\rangle-\frac{n}{r^2}E^{\rm aux}(p)-\frac 
{np}{r^2} \int \phi_p^4 \\ \nonumber && \!\!\!\!\! = 
\sum_{i=1}^n \int \left[ |\nabla_i f|^2 +\half \sum_{j,\, j\neq i}
v_a(|\x_i-\x_j|)|f|^2 
- 2p \ppr(\x_i^\perp)^2 |f|^2 \right] \prod_{k=1}^n 
\ppr(\x^\perp_k)^2 d^3
\x_k \ \!\! . \\
\eeqa
We now divide space in $\x^\perp$-direction into boxes of side length 
$s$, labeled by $\alpha$. Let $\phi_{\alpha,\rm max}$ and 
$\phi_{\alpha,\rm min}$ denote the maximal and minimal value of 
$\ppr$ inside box $\alpha$, respectively. We obtain
\beq\label{365}
E_{\rm N}^{\rm QM}(n,\ell,r,a)- \frac{n}{r^2}E^{\rm aux}(p)-\frac 
{np}{r^2} \int \phi_p^4 \geq \inf_{\{n_\alpha\}} \sum_\alpha \left[ 
E_\alpha(n_\alpha) - 2n_\alpha p  \phi_{\alpha,\rm max}^2\right] \ ,
\eeq
with 
\beq
E_\alpha(n)= \inf_{f}
\sum_{i=1}^n \frac{ \int_\alpha \left[ |\nabla_i f|^2 +\half 
\sum_{j,\, j\neq i}
v_a(|\x_i-\x_j|)|f|^2 \right] \prod_{k=1}^n \ppr(\x^\perp_k)^2 d^3
\x_k }{ \int_\alpha |f|^2 \prod_{k=1}^n \ppr(\x^\perp_k)^2 d^3
\x_k } \ .
\eeq
In (\ref{365}) the infimum is over all possible distributions of the 
$n$ particles into the boxes $\alpha$. As a first step we will 
replace $\phi_{\alpha,\rm max}^2$ by $\phi_{\alpha,\rm min}^2$ in the 
last term in (\ref{365}). The 
 error in doing so is bounded above by 
\beq
2n p \, \sup_\alpha\left( \phi_{\alpha,\rm max}^2-\phi_{\alpha,\rm 
min}^2\right)\leq 2\sqrt 2 n  \frac{ps}{r^3} \|\nabla^\perp 
\phi_p^2\|_\infty \ .
\eeq

To bound $E_\alpha$ from below, we divide $\alpha$ into even smaller 
boxes, denoted by $\beta$, with side length $t\leq s$, where $s/t\in 
\N$. Let $c_i$ denote the number of boxes with exactly $i$ particles. 
Then
\beq\label{367}
E_\alpha(n)\geq \inf_{\{c_i\}} \sum_{i\geq 0} c_i
\inf_{\beta\subset\alpha} E_\beta(i) \ .
\eeq
Note that the infimum is under the constraints $\sum c_i = (s/t)^2$ 
and $\sum c_i i = n$. For a lower bound on $E_\beta$, we use
\beq\label{rest}
E_\beta(n) \geq \left(\frac{\phi_{\beta,\rm min}^2}{\phi_{\beta,\rm 
max}^2}    \right)^n 
\inf_{f}
\sum_{i=1}^n \frac{ \int_\beta \left[ |\nabla_i f|^2 +\half 
\sum_{j,\, j\neq i}
v_a(|\x_i-\x_j|)|f|^2 \right] \prod_{k=1}^n  d^3
\x_k }{ \int_\beta |f|^2 \prod_{k=1}^n d^3
\x_k } \ .
\eeq
Fix some $\delta>0$, and assume that $\phi_{\alpha,\rm min}^2\geq 
\delta/r^2 $. Then
\beq
\left(\frac{\phi_{\beta,\rm min}^2}{\phi_{\beta,\rm 
max}^2}\right)^n\geq 1-\sqrt2 n\frac{t}{r}\frac{\|\nabla^\perp 
\phi_p^2\|_\infty }{\delta} 
\eeq
(compare with (\ref{334})). The rest of (\ref{rest}) can be bounded 
below by the same method as in the previous subsection. The only 
difference lies in the fact that $b_r$ is replaced by the constant 
function, and the integrations are only over the box $\beta$. The 
result is (compare with (\ref{lbthm})) 
\beqa \nonumber 
\!\!\!\!\!\!\!\! E_\beta(n) &\!\!\! \geq& \!\!\! E^{\rm 1D}_{\rm 
N}(n,\ell, 8\pi a/t^2 ) \\ \label{370}
&&\!\!\! \cdot\left(1-C
n\left(\frac{a}t\right)^{1/8}\left[1+\frac{n
t}{\ell}\left(\frac{a}t\right)^{1/8}\right]\right)\left(
1-\sqrt2n\frac{t}{r}\frac{\|\nabla \phi_p^2\|_\infty }{\delta} \right)
\ \!\! .
\eeqa
To proceed we need an 
 explicit lower bound on $E^{\rm 1D}_{\rm N}$ that will be proved in 
 the Appendix.

\begin{lem}\label{L2}
There is a finite number $C>0$ such that 
\beq\label{L2eq}
 E_{\rm N}^{\rm 1D}(n,\ell,g)\geq
\frac 12 \frac {n(n-1)}\ell g \left(1-C n (\ell g)^{1/2}\right) 
\ .
\eeq
\end{lem}

Applying this lemma to (\ref{370}), we obtain
\beqa\nonumber
E_\beta(n) &\geq&  \frac{n(n-1)}\ell \frac{4\pi a}{t^2}
\left(1-C
n\left(\frac{a}t\right)^{1/8}\left[1+\frac{n
t}{\ell}\left(\frac{a}t\right)^{1/8}\right]\right)\\ 
\label{370b}&&\cdot\left(
1-\sqrt2n\frac{t}{r}\frac{\|\nabla^\perp \phi_p^2\|_\infty }{\delta} 
\right)
\left(1-C n (\ell a/t^2)^{1/2}\right) \ .
\eeqa
Note that the right side is independent of $\beta\subset \alpha$. We
insert this bound in (\ref{367}), and use the following Lemma. It is a
simple generalization of a result of \cite{LY1998}. Although we need
at this point only the version proved in \cite{LY1998}, we state the
Lemma in this general form for later use. 
\begin{lem}[\cite{LY1998}]\label{L4}
    For $n\in\N\cup\{0\}$, let $E(n)$ be a sequence of non-negative real 
numbers that is superadditive, i.e., $E(n_1+n_2)\geq E(n_1)+E(n_2)$, 
and bounded below by
\beq
E(n)\geq L(n) K(n) \ ,
\eeq
with $K:\R_+\to \R_+$ monotone decreasing, $L:\R_+\to \R_+$ convex, 
$L(0)=0$, and
\beq\label{lpr}
L'(x)\leq \frac{L(\lambda x)}{2\lambda x}
\eeq
for some $\lambda>1$ and all $x>0$. 
(Here $L'$ denotes the right derivative of $L$.) 
Let $c_n$ be a sequence of non-negative real numbers, with 
\beq
\sum_{n\geq 0} c_n \leq M \quad {\rm and\ } \sum_{n\geq 0} c_n n = N 
\ .
\eeq
Then
\beq
\sum_{n\geq 0} c_n E(n) \geq M L(N/M) K(\langle \lambda N/M\rangle) \ ,
\eeq
where $\langle x\rangle$ denotes the smallest integer $\geq x$. 
\end{lem}

The proof is given in the
Appendix.
Note that $\inf_\beta E_\beta(n)$ is certainly a superadditive 
function, as the infimum over superadditive functions. Therefore we 
can apply Lemma~\ref{L4} with $L(x)=x[x-1]_+$ and 
$\lambda=4$, to 
(\ref{367}), together with (\ref{370b}), and obtain
\beq\label{370c}
E_\alpha(n_\alpha) \geq \frac{4\pi a n_\alpha^2}{\ell 
s^2}\left(1-\frac 1{n_\alpha} \frac{s^2}{t^2}\right){\cal 
R}(n_\alpha) 
\ ,
\eeq
where
\beqa\nonumber
\!\!\!\!\!\!\!\!\!\!{\cal R}(n)\!\!&=&\!\!
\left(1-C \left\langle 4 
nt^2/s^2\right\rangle\left(\frac{a}t\right)^{1/8}
\left(1+\left\langle 4 nt^2/s^2\right\rangle\frac{t}{\ell } 
\left(\frac{a}{t}\right)^{1/8}
\right)\right) \\ 
&&\!\! \cdot
\left( 1-\sqrt2 \left\langle 4 nt^2/s^2\right\rangle\frac{t}{r 
}\frac{\|\nabla^\perp \phi_p^2\|_\infty }{\delta} \right)\left(1-C 
\left\langle 4 nt^2/s^2\right\rangle (\ell a/t^2)^{1/2}\right) \ \! .
\eeqa
We now insert this bound in (\ref{365}), and use $n_\alpha\leq n$ in 
the error terms. Note that ${\cal R}$ is monotone decreasing in $n$, 
hence ${\cal R}(n_\alpha)\geq {\cal R}(n)$. Minimizing over 
$n_\alpha$ gives
\beq
E_\alpha(n_\alpha) - 2n_\alpha p  \phi_{\alpha,\rm min}^2
\geq -\frac{\ell s^2 p^2 \phi_{\alpha,\rm min}^4}{4\pi a}
\left(1+\frac{2\pi a r^2}{\ell t^2 p \delta}\right)^2 
\frac 1 { {\cal R}(n)} \ .
\eeq
This holds for boxes $\alpha$ where $\phi_{\alpha,\rm min}^2\geq 
\delta/r^2$. In boxes where this is not the case, we simply use 
positivity of $E_\alpha$ (which follows from positivity of $v_a$) in 
the form
\beq
E_\alpha(n_\alpha) - 2n_\alpha p  \phi_{\alpha,\rm min}^2
\geq - 2 n_\alpha \frac{p \delta}{r^2} \ .
\eeq
Putting everything together, using $\sum_\alpha s^2 \phi_{\alpha,\rm 
min}^4\leq r^{-2} \int \phi_p^4$ and choosing
\beq
p=\frac {4\pi a n}{\ell} \ ,
\eeq
we obtain
\beqa\nonumber
&&\!\!\!\!E^{\rm QM}_{\rm N}(n,\ell,r,a)\geq \frac n{r^2} E^{\rm 
aux}(4\pi an/\ell) 
\\&&\!\!\!\!- \frac{4\pi a n^2}{\ell r^2}\left( \sqrt 8 
\|\nabla^\perp \phi_p^2\|_\infty \frac sr+ 2\delta + \int \phi_p^4 
\left[\left(1+\frac{r^2}{2t^2 \delta n}\right)^2\frac 1{{\cal 
R}(n)}-1\right]\right) .
\eeqa
We are still free to choose the parameters $t$, $s$ and $\delta$.

It remains to derive a lower bound on $E^{\rm aux}$. This will be done
similarly to the lower bound on the 3D GP energy given in
Subsect.~\ref{3dgpsect}. Consider the auxiliary Schr\"odinger operator
\beq
H^{\rm aux}= -\Delta^\perp + V^\perp(\x^\perp) + 2p b(\x^\perp)^2 \ .
\eeq
Using $\phi_p$ as a trial function, we have
\beqa\nonumber
\infspec H^{\rm aux} &\leq& E^{\rm aux}(p) - p\int \phi_p^4 + 2p \int 
b^2 \phi_p^2 \\ &\leq& E^{\rm aux}(p) + p \int b^4 \ . \label{eq1}
\eeqa
On the other hand, using Temple's inequality (\ref{temple}),
\beqa\nonumber
\infspec H^{\rm aux} &\geq& 
e^\perp + 2p\int b^4 - \frac{4p^2 \int b^6}{\widetilde e^\perp - 2 p 
\int b^4} \\ &\geq & e^\perp + 2p \int b^4 \left(1-\frac{2p 
\|b\|_\infty^2}{\widetilde e^\perp - 2p \|b\|_\infty^2}\right) \ . 
\label{eq2}
\eeqa
Eqs.~(\ref{eq1}) and~(\ref{eq2}) together give
\beq
E^{\rm aux}(p)\geq e^\perp + p \int b^4 \left(1-\frac{4p 
\|b\|_\infty^2}{\widetilde e^\perp - 2p \|b\|_\infty^2}\right) \ .
\eeq
We now choose
\beq
s=\eps r ,\quad t=\frac r{\eps \sqrt n} , \quad \delta= \eps \ ,
\eeq
and
\beq
\eps = n^{-1/14} + \left(\frac{n\ell a}{r^2}\right)^{1/8} +
\left( \left[1+ \frac{r}{\sqrt n \ell}\right] \left(\frac{\sqrt n 
a}{r}\right)^{1/4}\right)^{4/39}  \ ,
\eeq
and obtain as the final result
\beq\label{aabb}
E^{\rm QM}_{\rm N}(n,\ell,r,a)\geq \frac {ne^\perp}{r^2} +\frac{ n^2 
g}{2 \ell}\left(1- C \left[\eps + \frac{na}{\ell}\right]\right)  \ .
\eeq
Note that we did not pay any attention to the restriction $s/t\in \N$
in choosing $s$ and $t$. However, since, with our choice, $s/t=\eps^2
\sqrt{n}\geq n^{1/2-1/7}$, this can easily be made an integer by
replacing $\eps$ by some $\bar\eps$ with $\eps\leq \bar\eps\leq
2\eps$. This affects only the constant $C$ in (\ref{aabb}).

\subsection{Boundary conditions}\label{bcsect}

As a last step in this section, before giving the proof of our main
Theorem~\ref{T1}, we investigate the dependence of the ground state
energy of (\ref{13}) on the boundary conditions. In the upper bound
above we used Dirichlet boundary conditions for the energy $E^{\rm
1D}$, and Neumann boundary conditions for the lower bound. To relate
these energies to the energy with periodic boundary conditions and to
prove independence of boundary conditions in the thermodynamic limit,
we need the following Lemma.

\begin{lem}\label{L1}
Denote  $E_{\rm
p}^{\rm 1D}(n,\ell,g)$ the ground state energy of (\ref{13}) with
periodic boundary conditions, i.e., on the torus $[0,\ell]^n$.
Then there is a finite number $C>0$ such that 
\beq\label{first}
E_{\rm N}^{\rm 1D}(n,\ell,g)\leq E_{\rm p}^{\rm 1D}(n,\ell,g)\leq 
E_{\rm D}^{\rm 1D}(n,\ell,g) \ ,
\eeq
and
\beq\label{second}
E_{\rm D}^{\rm 1D}(n,\ell,g)\leq E_{\rm N}^{\rm 1D}(n,\ell,g)+ C 
\frac {n^{7/3}}{\ell^2} \ .
\eeq
\end{lem}

\begin{proof}
The first inequality (\ref{first}) is standard, noting that the 
interaction considered has zero range. For (\ref{second}) we use a 
result of \cite[Lemma~2.1.13 and Prop.~2.2.10]{robinson}, which 
implies, for $0<b<\ell/2$,
\beq
E_{\rm D}^{\rm 1D}(n,\ell+2b,g)\leq E_{\rm N}^{\rm 
1D}(n,\ell,g)+\frac {2n}{b^2} \ .
\eeq
Using $E_{\rm D}^{\rm 1D}(n,\ell+2b,g)=E_{\rm D}^{\rm 
1D}(n,\ell,g(1+2b/\ell))(1+2b/\ell)^{-2}\geq E_{\rm D}^{\rm 
1D}(n,\ell,g)(1+2b/\ell)^{-2}$ and $E_{\rm N}^{\rm 1D}(n,\ell,g)\leq 
\pi^2 n^3/(3\ell^2)$, we obtain
\beq
E_{\rm D}^{\rm 1D}(n,\ell,g)\leq E_{\rm N}^{\rm 1D}(n,\ell,g)+\frac 
{2n}{b^2}\left(1+2b/\ell\right)^2+\frac {n^3}{\ell^2}\frac{4\pi^2}3 
\frac b\ell\left( 1+b/\ell\right) \ .
\eeq
Now $b=\const \ell/n^{2/3}$ gives the desired result.
\end{proof}

\section{Proof of Theorem~\ref{T1}}\label{sect4}

With the results of the previous section in hand, we can now give the
proof of our main Theorem~\ref{T1}.  The proof will be divided into
four subsections, two for the upper and lower bound, respectively.  In
each subsection, we compare the ground state energy of $H_{N,L,r,a}$
with the ground state energy of one of the functionals in
Subsections~\ref{igss}--\ref{gtss}, which, as explained there and in
Subsection 2.6, is asymptotically equal to $E(N,L,g)$ in the
respective parameter region.  Combining all the bounds obtained, this
will prove Theorem~\ref{T1}, together with the claimed uniformity in
the parameters.  The corresponding convergence of the ground state
particle density, as stated in Theorems~\ref{T2.1}--\ref{T2.5},
follows in a standard way by variation with respect to the external
potential $V_L(z)$ (compare with Props.\ 2.1 and 2.2 in
Subsect.~\ref{sspr}).  Since the proof of the energy convergence is
already quite lengthy by itself, the simple modifications necessary
for a proof of the density convergence will be omitted.

Let again $\bar L=N/\bar\rho$ denote the extension of the system in
$z$-direction. As already mentioned in the beginning of
Section~\ref{finsect}, it will be necessary, for the lower bound to
$E^{\rm QM}$, to consider the case of small and large $Nr/\bar L$
separately. We will divide space in $z$-directions into small boxes,
and use the bounds of Section~\ref{finsect} in each box. To control
the number of particles in each box, Lemma~\ref{L4} will be
essential. For small $Nr/\bar L$, where $r$ is smaller than the mean
particle distance, we will use the lower bound given in
Thm.~\ref{finthm}. For larger values of $Nr/\bar L$, where $r$ is
actually bigger than the mean particle distance, the lower bound of
Thm.~\ref{finthm2} will be used instead. Note that this distinction is
only relevant in Regions 1--3, since in Regions~4 and~5 $Nr/\bar L=
r\bar\rho\ll 1$ by condition (\ref{condition}). Hence $r$ is always
smaller than the mean particle distance in Regions 4 and 5.

\subsection{Upper bound for Regions 1--3}\label{ssu13}

For an upper bound that gives the right asymptotics as long as
$g/\bar\rho\ll 1$, we can essentially use the same technique as in
\cite{lsy1}. The results of Section~\ref{finsect} are not needed in 
this case. As a trial function, we use
\beq
\Psi(\x_1,\dots,\x_N)=F(\x_1,\dots,\x_N) \prod_{k=1}^N 
b_r(\x_k^\perp) \phi^{\rm GP}(z_k) \ ,
\eeq
where $\phi^{\rm GP}=(\rho^{\rm GP}_{N,L,g})^{1/2}$
(cf. Subsect.~\ref{gpss}), and $F$ is the `Dyson wave function', 
described
in \cite{dyson,lsy1}. The result is
\beq
E^{\rm QM}(N,L,r,a)\leq \frac{Ne^\perp}{r^2}+ E^{\rm 
GP}(N,L,g)\left(1+ C a \|b_r\|_\infty^{2/3} \|\phi^{\rm 
GP}\|_\infty^{2/3}\right) \ ,
\eeq
as long as $ a \|b_r\|_\infty^{2/3} \|\phi^{\rm 
GP}\|_\infty^{2/3}\leq 
1$. Note that, by the same proof as in Lemma 2.1 of \cite{lsy2}, $g 
\|\phi^{\rm GP}\|_\infty^{2}\leq 2E^{\rm GP}/N$, and therefore
\beq
E^{\rm QM}(N,L,r,a)\leq \frac{Ne^\perp}{r^2}+ E^{\rm 
GP}(N,L,g)\left(1+ \const a^{2/3}N^{-1/3} E^{\rm GP}(N,L,g)^{1/3} 
\right) \ .
\eeq
Now, in Regions 1--3, $E^{\rm GP}(N,L,g)\sim E(N,L,g)$, and
\beq
a^2 E^{\rm GP}(N,L,g)/N \sim a^2 \left(L^{-2}+ g \bar\rho\right) \sim 
(a/L)^2+  \frac g{\bar\rho} 
\left(g\bar\rho r^2\right)^2 \ll 1 \ .
\eeq

\subsection{Upper bound for Regions 4 and 5}\label{sectup4}

If $g/\bar\rho$ is not small, the method of the previous subsection
does not work, and we have to proceed differently. As in
Subsect.~\ref{bcsect}, let $E_{\rm p}^{\rm 1D}(n,\ell,g)$ denote the
ground state energy of (\ref{13}) on an interval of length $\ell$,
with periodic boundary conditions, and write $E_{\rm p}^{\rm
1D}(n,\ell,g)=n^3 e_n(g\ell/n)/\ell^2$. In
\cite{LL} it is shown that, for every fixed $t\geq 
0$,
$\lim_{n\to\infty} e_n(t)= e(t)$. Since the functions are
monotone increasing, concave and bounded in $t$, the convergence
is actually uniform in $t$. By Lemma~\ref{L1} of 
 Subsection 3.5, the same is true
with $E_{\rm D}^{\rm 1D}$ instead of $E_{\rm p}^{\rm 1D}$. Hence we
get the estimate
\beq\label{thermo}
E_{\rm D}^{\rm 1D}(n,\ell,g)\leq \frac{n^3}{\ell^2} \Big( e(g\ell/n) +
\delta(n) \Big)
\eeq
for some bounded function $\delta$ satisfying
$\lim_{n\to\infty}\delta(n)=0$. Without loss of generality we may
assume that $\delta$ is monotone decreasing.

Let $\rho$ be the minimizer of the LL functional (\ref{llfunct}) under
the normalization condition $\int \rho = N$. Note that $\rho$ has
compact support, with radius $R=L(\mu L^2)^{1/s}$, where 
$\mu=\partial E^{\rm LL}(N,L,g)/\partial N$. (This $R$ is 
different from that in Eq.\ (\ref{Rin3.1}).)  By 
monotonicity and concavity of $E^{\rm LL}$ in $g$, and by the scaling 
relation (\ref{scalll}), 
\beq\label{mumu}
\frac {E^{\rm LL}(N,L,g)}N \leq \mu \leq 3 \frac {E^{\rm LL}(N,L,g)}N 
\ .
\eeq

Divide $\R$ in $z$-direction into intervals of length $\ell$, labeled
by $\alpha$, with $R_0<\ell<R/2$. Let $n_\alpha\in \N$ be a 
collection of integers such that $\sum_\alpha n_\alpha = N$.
Let $V_\alpha=
\sup_{z\in\alpha} \V_L(z)$, and denote $E^{\rm QM}_{\rm
D}(n,\ell,r,a)$ the ground state energy of (\ref{defh}) in a box of
side length $\ell$ and with Dirichlet boundary conditions. By
confining the particles into different boxes of length $\ell -R_0$, a 
distance $R_0$ apart, we get the estimate
\beq
E^{\rm QM}(N,L,r,a)\leq \sum_\alpha \left[ E^{\rm QM}_{\rm 
D}(n_\alpha,\ell-R_0,r,a) + V_\alpha n_\alpha\right] \ .
\eeq 
Using (\ref{ubthm}) and (\ref{thermo}) as well as monotonicity of
$e$, we obtain
\beq\label{aobt}
E^{\rm QM}(N,L,r,a)-\frac {Ne^\perp}{r^2}
\leq \sum_\alpha\left[ \frac{n_\alpha^3}{\ell^2}\Big( e(g\ell/ 
n_\alpha)+\delta( n_\alpha)\Big){\cal R}(n_\alpha)
 + V_\alpha n_\alpha \right] \ , 
\eeq
with
\beq
{\cal R}(n)= \frac 1{(1-R_0/\ell)^2}
\left(1+ C \left[\left(\frac{n a}{r}\right)^2(1+
g\ell)\right]^{1/3}\right) \ ,
\eeq
provided we choose $n_\alpha$ and $\ell$ such that the term in square
brackets is less than 1. Note the additional factor 
$(1-R_0/\ell)^{-2}$,
which is due to the fact that the size of the box is only $\ell
-R_0$. Now let $\bar n_\alpha=\int_\alpha
\rho(z)dz$, and $n_\alpha=\langle
\bar n_\alpha\rangle$,
where $\langle x\rangle $ denotes the smallest integer $\geq x$. With
this choice $\sum_\alpha n_\alpha \geq N$, but by monotonicity of
(\ref{aobt}) in $N$ we can plug in these values of $n_\alpha$ for an
upper bound.

Since $n_\alpha\leq \bar n_\alpha+ 1$, and $e$ and $\delta$ are 
monotone increasing and decreasing, respectively, we obtain
\beq\label{aobt2}
(\ref{aobt})\leq \sum_\alpha\left[ \frac{ \bar 
n_\alpha^3}{\ell^2}\Big( e(g\ell/\bar n_\alpha)+\delta(\bar 
n_\alpha)\Big)\left(1+\frac 1{\bar n_\alpha}\right)^3{\cal R}(
\ell\|\rho\|_\infty+1)
+ V_\alpha ( \bar n_\alpha+1) \right] \ .  
\eeq
Here we estimated $\bar n_\alpha$ by $\ell \|\rho\|_\infty$ in ${\cal R}$. 
Denote $\hat V_\alpha=\min_{z\in \alpha} V_L(z)$. We estimate
$V_\alpha\leq \hat V_\alpha + \const L^{-2} (\ell/L) (R/L)^{s-1}$ in
boxes where $n_\alpha>0$. Using $(R/L)^s\leq 3 L^2 E^{\rm
LL}(N,L,g)/N$ (see (\ref{mumu})), we therefore see that the error in
replacing $V_\alpha$ by $\hat V_\alpha$ is, in total, bounded above by
$\const E^{\rm LL}(N,L,g) (\ell/R)$.

Fix some $0<\eps<1 $. We first bound the contribution to (\ref{aobt2})
from boxes where $\bar n_\alpha\leq 1/\eps$. Now both $e$ and 
$\delta$ are
bounded, the number of boxes with nonzero $\bar n_\alpha$ is bounded
by $(R/\ell)+2$, and $\hat V_\alpha \leq L^{-2} (R/L)^s$ in these
boxes. Therefore this contribution is bounded above by
\beq\label{49}
\const \frac {R}{\eps^2 \ell}\left(\frac 1{\eps \ell^2}
{\cal R}(\ell\|\rho\|_\infty + 1 )
+\eps \frac{E^{\rm LL}(N,L,g)}N\right) \ \! .
\eeq
For the remaining boxes, we use $\bar n_\alpha\geq 1/\eps$ and $\bar 
n_\alpha\leq \ell \|\rho\|_\infty$ to obtain
\beqa\nonumber
(\ref{aobt2})\!\! &\leq&\!\! (\ref{49}) +\const E^{\rm LL}(N,L,g) 
\frac \ell R 
\\
\nonumber &&\!\!+\, (1+\eps)\cdot \sum_\alpha\left[ \frac{ 
n_\alpha^3}{\ell^2}\Big( 
e(g\ell/n_\alpha)+\delta(1/\eps)\Big)(1+\eps)^2 {\cal 
R}(\ell\|\rho\|_\infty + 1)
 + \hat V_\alpha n_\alpha \right] \ . \\
\eeqa
Since $x\mapsto x^3 e(1/x)$ is convex, we can use Jensen's inequality 
to bound the sum from above by
\beq
\int_\R  \left[ \rho(z)^3\Big( e(g/\rho(z)) 
+\delta(1/\eps)\Big)\left(1+\eps\right)^2{\cal R}(\ell \|\rho\|_\infty
+ 1)
 + V(z)\rho(z)\right] dz \ .
\eeq
Now, by the scaling 
(\ref{rhoscaling}), $\rho(z)=\gamma\tilde \rho_{g/\gamma}(\gamma z/N)$, where 
$\gamma= (N/L)N^{-2/(s+2)}$ and
$\tilde\rho_{g/\gamma}$ is a function that depends only on 
$g/\gamma$. Therefore
$\|\rho\|_\infty \leq \gamma \|\tilde\rho_{g/\gamma}\|_\infty$, and 
$\|\rho\|_3^3 \leq N \gamma^2 \| \tilde\rho_{g/\gamma}\|_\infty^2$. 

We choose, for some $0<\hat \eps<\eps$, $\ell=(\hat\eps \gamma)^{-1}$,
and use
\beq\label{rng}
E^{\rm LL}(1,1,g/\gamma)^{1/s}\leq \frac{\gamma R}N \leq \pi^{2/s} \ .
\eeq
Putting everything together, we get, for $C$ denoting some universal 
constant, 
\beqa\nonumber
&& E^{\rm QM}(N,L,r,a)-\frac {Ne^\perp}{r^2}\\ &&\leq \Big(
E^{\rm LL}(N,L,g)+N\gamma^2 \left( \|\tilde\rho_{g/\gamma}\|_\infty^2 
\delta(1/\eps) + C (\hat\eps/\eps)^3\right)\Big)\cdot {\cal R}' \ , 
\label{415}
\eeqa
where
\beqa\nonumber
&&{\cal R}'
=\left(1+\frac C{N \hat \eps E^{\rm 
LL}(1,1,g/\gamma)^{1/s}}+C 
\frac{\hat\eps}\eps\right)\left(1+\eps\right)^3 \\  &&\qquad\cdot 
\left(1-\frac
{a R_0}{r^2}\hat \eps \frac\gamma g\right)^{-2} \left(1+ C 
\left(\frac{a
\|\tilde \rho_{g/\gamma}\|_\infty}{\hat \eps r}\right)^{2/3}\left(1+
\frac g{\hat\eps\gamma}\right)^{1/3}\right) \ .
\eeqa
The choice of $\eps$ and $\hat \eps$ is determined by $a/r$, 
$g/\gamma$ and $N$. The bound is
uniform in $g/\gamma$ for bounded $g/\gamma$ and $\gamma/g$.

If $g/\gamma\to \infty$ as $N\to\infty$ (Region 5), we can use the 
same method to obtain an upper bound, replacing the bound 
(\ref{ubthm}) by (\ref{uppbou2})
in (\ref{aobt}). This gives a  bound uniform in $g/\gamma$ (for 
$\gamma/g$ 
bounded). Combined with the result (\ref{415}), this shows that in 
the limit $N\to\infty$ and $r/L\to 0$
\begin{equation}
\limsup \frac {E^{\rm QM}(N,L,r,a)-Ne^\perp/r^2}{E^{\rm 
LL}(N,L,g)}\leq 1\ ,
\end{equation}
uniformly in the parameters, as long as $a/r\to 0$ and $\gamma/g$ 
stays
bounded.

\subsection{Lower bound for Regions 3--5}\label{subsec43}

We now derive a lower bound on $E^{\rm QM}$ that will give the right
asymptotics in Regions 3--5. As in the upper bound, given in
Subsection~\ref{sectup4}, we will use the box method, this time with
Neumann boundary conditions. In each box, the results of
Section~\ref{finsect} will be used. In analogy to (\ref{thermo}) we
infer from \cite{LL} and Lemma~\ref{L1} that
\beq\label{thermo2}
E^{\rm 1D}_{\rm N}(n,\ell,g)\geq \frac{n^3}{\ell^2} \Big( e(g\ell/n) 
- 
\delta(n) \Big)
\eeq
for some bounded, monotone decreasing function $\delta$ satisfying
$\lim_{n\to\infty}\delta(n)=0$. This will be used in the bound for
Regions 4 and 5. If $g\ell/n$ is small, however, we use 
$e(g\ell/n)\leq
\half g\ell/n$ and (\ref{L2eq}) to obtain
\beq\label{thermo3}
E_{\rm N}^{\rm 1D}(n,\ell,g)\geq \frac{n^2(n-1)}{\ell^2} 
e(g\ell/n)\Big(1 - 
C n
(g\ell)^{1/2}\Big) \ .
\eeq

We divide $\R$ in $z$-direction into intervals of side length $M$,
labeled by $\alpha$. Denote $E^{\rm QM}_{\rm N}(n,M,r,a)$ the ground
state energy of (\ref{defh}) in a box of side length $M$ and with
Neumann boundary conditions. Let again $\hat V_\alpha=
\inf_{z\in\alpha} V_L(z)$, and $V_\alpha=
\sup_{z\in\alpha} V_L(z)$.  By confining the particles into different
boxes and neglecting the interaction between different boxes, we get
the estimate 
\beq\label{esti} 
E^{\rm QM}(N,L,r,a)\geq
\inf_{\{N_\alpha\}}{\sum_\alpha} \left[ E^{\rm QM}_{\rm
    N}(N_\alpha,M,r,a) + \hat V_\alpha N_\alpha\right] \ , 
\eeq 
where the infimum is over all distributions of the $N$ particles among
the boxes $\alpha$.  As in the upper bound, we can estimate the
difference of the maximum and minimum of $V_L$ for boxes alpha
$\alpha$ inside some interval $[-R,R]$ by $ \const L^{-2} (M/L)
(R/L)^{s-1}$. For boxes outside $[-R,R]$ we use $\hat V_\alpha\geq
V_\alpha(1-s M/R)$.  Choosing $R$ the radius of the LL minimizer, and
$M=\eps R$ for some $0<\eps<1$, we see that, analogously to the upper
bound, the error in replacing $\hat V_\alpha$ by $ V_\alpha(1-s \eps)$
is, in total, bounded above by $\const \eps E^{\rm LL}(N,L,g)$.

We now have to estimate $E^{\rm QM}_{\rm N}(N_\alpha,M,r,a)$ from
below. We cannot directly use (\ref{lbthm}), because this bound is
not uniform in the particle number. Instead we proceed similarly to
\cite{LY1998}. We divide the interval $M$ again into smaller intervals
of length $\ell=\eta M$, where $1/\eta\in\N$.  Neglecting the
interaction between different boxes, we obtain
\beq\label{reads}
E^{\rm QM}_{\rm N}(N_\alpha,M,r,a)\geq \inf_{\{c_n\}}
\sum_{n=1}^{N_\alpha} c_n E^{\rm QM}_{\rm N}(n,\ell,r,a) \ ,
\eeq
where $c_n$ denotes the number of boxes containing exactly $n$
particles, and the infimum is over all $c_n\in \N$ under the condition
$\sum_n c_n n=N_\alpha$ and $\sum_n c_n = M/\ell=\eta^{-1}$.

Fix some $0<\chi<1$, and consider the case $n\geq 1/\chi$. 
Then, using (\ref{lbthm}) and (\ref{thermo2}),
\beq\label{still}
E^{\rm QM}_{\rm N}(n,\ell,r,a)-\frac{ne^\perp}{r^2} 
\geq \frac{n^3}{\ell ^2} e(g\ell /n)
\Big(1-\nu(n)\Big) \ ,
\eeq
with
\beq
\nu(n)= \frac{\delta(1/\chi)}{e(g\ell/n)}+  C  n \left(\frac{ 
a}{r}\right)^{1/8}\left[1+\frac{nr}{\ell}\left(\frac{a}{r}\right)^{1/8}\right]  
\ .
\eeq
Note that $\nu(n)$ is monotone increasing in $n$. We now use
Lemma~\ref{L4} from Subsection 3.4, with $L(x)=x^3e(g\ell/x)$. Note
that for this $L$ (\ref{lpr}) holds with $\lambda=6$, since $e$ is a
monotone increasing and concave function, with $e(0)=0$. Let
$N'=\sum_{n\geq 1/\chi}c_n n$. The contribution from $n\geq 1/\chi$ to
the sum in (\ref{reads}) will be bounded below using Lemma~\ref{L4}
and (\ref{still}). For $n<1/\chi$, we simply use $E^{\rm QM}_{\rm
N}(n,\ell,r,a)\geq n e^\perp/r^2$. We thus obtain
\beq\label{obta}
E_{\rm N}^{\rm QM}(N_\alpha,M,r,a)-\frac {N_\alpha e^\perp}{r^2}\geq 
\frac
{N'^3}{M^2} e(gM/N')  \Big(1- \nu(\langle 6 N'\eta\rangle) \Big) \ .
\eeq
Using $N_\alpha\geq N'\geq N_\alpha-1/(\eta \chi)$, this gives
\beq
E_{\rm N}^{\rm QM}(N_\alpha,M,r,a)-\frac {N_\alpha e^\perp}{r^2}\geq 
\frac
{N_\alpha^3}{M^2} e(gM/N_\alpha) \left(1-\frac 1{N_\alpha\eta 
\chi}\right)^3
\Big(1- \nu(\langle 6 N_\alpha\eta\rangle) \Big) \ .
\eeq
Now if $\hat\eps N_\alpha\geq 2$ for some $0<\hat\eps<\chi$, we can 
choose
$\half\hat\eps\leq\delta\leq\hat\eps$ such that $\delta N_\alpha\in 
\N$, and
take $\eta=(\delta N_\alpha)^{-1}$. We also use $\langle 
6/\delta\rangle\leq
\langle 12/\hat\eps\rangle \leq 13/\hat\eps$ (since $\hat\eps<1$).
Note that, using (\ref{rng}) and $\delta\leq \hat\eps$,
\beq
\frac{\hat \eps}{13} g\ell \geq \frac \eps {13} \frac g\gamma E^{\rm 
LL}(1,1,g/\gamma)^{1/s}\equiv \eps \xi(g/\gamma) \ .
\eeq
Therefore
\beq\label{obta2} 
E_{\rm N}^{\rm QM}(N_\alpha,M,r,a)-\frac {N_\alpha e^\perp}{r^2} 
\geq \frac
{N_\alpha^3}{M^2} e(gM/N_\alpha) {\cal R} \ ,
\eeq
with
\beq
{\cal R}=  \left(1-\frac {\hat \eps}\chi\right)^3
\left(1 -\frac{ \delta(1/\chi)}{e(\eps\xi(g/\gamma))} 
-\frac C{  \hat\eps} \left(\frac{a}{r}\right)^{1/8}\left[1+\frac 
1{\eps\xi(g/\gamma)} \frac ar  \left(\frac{a}{r}\right)^{1/8} \right] 
\right) \ .
\eeq
Here we used (\ref{rng}) and $N_\alpha\leq N$ in the last error term.

The bound (\ref{obta2}) holds for $\hat\eps N_\alpha\geq 2$. If
$N_\alpha< 2/\hat\eps$, however, we use
\beq\label{naa}
\frac{N_\alpha^3}{M^2}e(gM/N_\alpha)\leq N_\alpha \frac{4\pi^2}{3 
\hat\eps^2 M^2} \ .
\eeq
Using these bounds in (\ref{esti}), we obtain
\beqa\nonumber
&&E^{\rm QM}(N,L,r,a)-\frac {N e^\perp}{r^2}+ C\eps E^{\rm 
LL}(N,L,g)+N\frac{4\pi^2}{3 \hat\eps^2 M^2}  
\\ \label{square} &&\geq \inf_{\{N_\alpha\}}{\sum_\alpha}
\left[
\frac {N_\alpha ^3}{M^2} e(gM/N_\alpha) + V_\alpha  N_\alpha \right] 
{\cal R}(1-s\eps) \ .
\eeqa
Note that, by (\ref{rng}) and (\ref{scalll}), 
\beq
\frac {N}{M^2}=\frac N{\eps^2 R^2}\leq E^{\rm LL}(N,L,g)\frac 
1{\eps^2N^2 E^{\rm LL}(1,1,g/\gamma)^{1+2/s}} \ .
\eeq
Define $\rho(z)=\sum_\alpha N_\alpha \chi_\alpha(z)$, where 
$\chi_\alpha$ is the characteristic function of the interval 
$\alpha$. The sum in (\ref{square}) is bounded 
below by $\E^{\rm LL}[\rho]\geq E^{\rm LL}(N,L,g)$, and therefore 
\beq
E^{\rm QM}(N,L,r,a)-\frac {N e^\perp}{r^2}\geq E^{\rm LL}(N,L,g) 
 {\cal R}\left(1 - C\eps-\frac {4\pi^2}{3\eps^2\hat\eps^2N^2 E^{\rm 
LL}(1,1,g/\gamma)^{1+2/s}}\right) \ \!\! . \label{aaa}
\eeq
The choice of $\eps$, $\hat\eps$ and $\chi$ is determined by
$g/\gamma$ and $a/r$. They can be chosen such that (\ref{aaa}) gives
the correct lower bound in the limit considered, uniformly in
$g/\gamma$ for bounded $\gamma/g$. This finishes the proof of the
lower bound for Regions 4 and 5.
\medskip

If $g/\gamma\to 0$ as $N\to \infty$, we can use exactly the same 
strategy, with (\ref{thermo3}) replacing the bound (\ref{thermo2}). 
Considering the case $n\geq 1/\chi$, (\ref{still}) is still valid, 
but with $\nu(n)$ replaced by
\beq\label{nupr}
\nu'(n)= \chi+ \const n \sqrt{ g\ell} +  \const  n 
\left(\frac{a}{r}\right)^{1/8}\left[1+\frac{nr}{\ell}\left(\frac{a}{r}\right)^{1/8}\right]  
\ .
\eeq
For $N_\alpha\geq \max\{2/\hat\eps, \eps^2 N\}$ we proceed exactly as
above. (We recall that $0<\eps<1$, and $M=\eps R$.) The reason why we 
have to ensure that $N_\alpha\geq \eps^2 N$
is the second term in (\ref{nupr}), where we want $\ell$ to be
small. (Note that we choose $\ell=M/(\delta N_\alpha)$.) For 
$N_\alpha<
\max\{2/\hat\eps, \eps^2 N\}$, we replace the bound (\ref{naa}) by
\beqa\nonumber
\!\!\!\!\!\!\!\!\!\! \frac{N_\alpha^3}{M^2}e(gM/N_\alpha)&\leq& 
\frac{N_\alpha}{\eps R}\frac g2 \max\{2/\hat\eps, \eps^2 N\} \\ 
&\leq&  \frac \eps 2\frac{N_\alpha}{N}  E^{\rm 
LL}(N,L,g)\left(1+\frac{2}{N\eps^2\hat\eps}\right) 
\frac{g/\gamma}{E^{\rm LL}(1,1,g/\gamma)^{1+1/s}} \ \! ,
\eeqa
where we used $e(x)\leq \half x$ and (\ref{rng}). Note that, for small
$g/\gamma$, the last fraction is order 1. We obtain
\beqa \label{aaa2}
E^{\rm QM}(N,L,r,a)-\frac {N e^\perp}{r^2} \geq  E^{\rm
LL}(N,L,g) {\cal R'}\  , 
\eeqa
with ${\cal R}'$ given by  
\beqa\nonumber
{\cal R'}&=&\left(1-\frac {\hat \eps}\chi\right)^3
\left(1 -\chi - C\frac 1{\hat\eps} \sqrt{\frac{ g}{\eps \hat 
\eps\gamma}}
-\frac C{  \hat\eps} \left(\frac{a}{r}\right)^{1/8}\left[1+\frac 
1{\eps\xi(g/\gamma)} \frac ar  \left(\frac{a}{r}\right)^{1/8} \right] 
\right) \\ &&\cdot \left(1 - C\eps- \frac \eps
2\left(1+\frac{2}{N\eps^2\hat\eps}\right) \frac{g/\gamma}{E^{\rm
LL}(1,1,g/\gamma)^{1+1/s}} \right) \ .
\eeqa
Here we used again (\ref{rng}) to estimate $R$ from above, and 
$\eps^2 N\leq N_\alpha\leq N$ in the error term. 
This proves the lower bound in Region 3, as long as 
$a/(r\xi(g/\gamma))$ stays bounded (or at least does not increase too 
fast as $a/r$ goes to zero with $N$). Note that, for small $g/\gamma$,
\beq
\frac ar \frac 1{\xi(g/\gamma)} \sim  \frac a r 
(\gamma/g)^{(s+2)/(s+1)} \sim \frac {rN}{\bar L_{\rm TF}} \ , 
\eeq
with $\bar L_{\rm TF}$ defined in (\ref{scaleddens}), 
and hence the bound is uniform for $rN/\bar L_{\rm TF}$ bounded.

As explained briefly in the introduction to this section, if $Nr/\bar
L_{\rm TF}$ is not small, we have to use Thm.~\ref{finthm2} instead of
Thm.~\ref{finthm}. Hence we will now assume that $A\equiv (rN/\bar 
L_{\rm
TF})^{1/3}\geq 1$, but still $g/\gamma\to 0$ as $N\to\infty$. Instead
of (\ref{lbthm}) we will use the bound (\ref{final}) in
(\ref{still}). This gives (\ref{still}) with $\nu(n)$ replaced by
\beq
\nu''(n) = C\left( \frac {na}{\ell} + \chi^{1/14} + \left(\frac{n\ell 
a}{r^2}\right)^{1/8} +\left( \left[1+ \frac{\sqrt\chi 
r}{\ell}\right]\left(\frac{\sqrt n a}{r}\right)^{1/4}\right]^{4/39} 
\right) \ .
\eeq
Let $0<\hat\eps<1$. For $N_\alpha\geq \max\{2 A^{2}/\hat\eps, \eps^2 
N\}$ we proceed as above, but choosing $\eta =A^{2}/(\delta 
N_\alpha)$ for $\half\hat\eps\leq \delta\leq \hat\eps$. Moreover, we
choose $\chi=1/(\widetilde \eps N_\alpha \eta)$. Using $R\sim \bar 
L_{\rm TF}$, this gives 
(\ref{aaa2}), with ${\cal R}'$ replaced by
\beqa\nonumber
{\cal R}'' \!\!\!&=\!\!\!& 
\left(1-C\left[\frac{Na}{\eps R}+ \left(\frac {\hat\eps}{\widetilde 
\eps A^{2}}\right)^{\frac 1{14}} + \left(\frac {a 
A}{r\eps\hat\eps^2}\right)^{\frac 18} + 
      \left( \left[1+\frac{\hat\eps^{1/2}}{\eps 
\widetilde\eps^{1/2}}\right]\left( \frac {a A}{\hat\eps 
r}\right)^{1/4}\right) ^{\frac 4{39}}  
      \right]\right) \\ && \cdot(1-\widetilde \eps)^3 \left(1 - 
C\eps- \frac \eps
      2\left(1+\frac{2 A^{2}}{N\eps^2\hat\eps}\right)
      \frac{g/\gamma}{E^{\rm LL}(1,1,g/\gamma)^{1+1/s}} \right) \ .
\eeqa
Note that, for $g/\gamma\ll 1$, $Na/R\sim a\bar\rho \ll 1$, and
$(aA/r)^3= (a/r)^2 Na/R$. Moreover, $A^{2}/N\ll N^{-1/3}$ if $\bar
L_{\rm TF}/L = (NgL)^{1/(s+1)}$ is bounded away from zero. Hence, for
bounded $1/A$, this gives the desired lower bound for Region 3.

In summary, we have thus shown that, in the limit $N\to\infty$ and 
$r/L\to 0$, 
\begin{equation}
\liminf \frac {E^{\rm QM}(N,L,r,a)-Ne^\perp/r^2}{E^{\rm 
LL}(N,L,g)}\geq 1\ ,
\end{equation}
uniformly in the parameters, provided (\ref{condition}) holds, $a/r\to
0$ and $1/(NgL)$ stays bounded. This finishes the proof of the lower
bound for Regions 3--5.

\subsection{Lower bound for Regions 1 and 2}\label{ssl12}

We are left with the lower bound for Regions 1 and 2. 
We proceed similarly to \cite[Sect.~5.1]{LSSY}. Let 
$\rho_{N,L,g}^{\rm GP}$ 
denote the minimizer of the GP functional (\ref{GPfunct}) under the 
normalization condition $\int_\R \rho=N$, and let 
$\phi(z)=(\rho_{N,L,g}^{\rm 
GP}(z))^{1/2}$. We write a general wave function $\Psi$ as 
\beq
\Psi(\x_1,\dots,\x_N)= F(\x_1,\dots,\x_N) \prod_{k=1}^N \phi(z_k) 
b_r(\x^\perp_k) 
\eeq
and assume that $\langle\Psi|\Psi\rangle=1$. In evaluating the 
expectation value of $H_{N,L,r,a}$, we use partial integration and 
the GP equation
\beq
-\phi''+V\phi + g|\phi|^2 \phi =\left(E^{\rm GP}(N,L,g) + \frac g2 
\int |\phi|^4\right) \phi \ .
\eeq
Moreover, we split a fraction of the kinetic energy into a part where
the particles are closer than a distance $T>R_0$, and it's
complement. This splitting will be important in the proof of BEC
below. More precisely, for fixed $i$ and $\x_j$, $j\neq i$, let
\beq\label{defchi}
\chi_{i,T}(\x) = \left\{ \begin{array}{ll} 1 & {\rm if\ }\min_{k,\,
      k\neq i} |\x-\x_k|\geq T \\ 0 & {\rm otherwise} \ ,
  \end{array}\right. 
\eeq 
and let $\bar\chi_{i,T}= 1 - \chi_{i,T}$.  Then, for $0\leq\eps\leq 
1$, 
\beqa\nonumber
\langle\Psi|H_{N,L,r,a}|\Psi\rangle\!&=&\!
\frac{Ne^\perp}{r^2} + E^{\rm GP}(N,L,g)+ \frac g2 \int |\phi|^4+ 
Q(F) \\
 &&\! + (1-\eps)
\int |\nabla_i
      F|^2\chi_{i,T}(\x_i) \prod_{k=1}^N \phi(z_k)^2
  b_r(\x^\perp_k)^2 d^3\x_k \ \! ,
\eeqa
with
\beqa\nonumber
Q(F)\!\!&=&\!\!
\int \left(\sum_{i=1}^N \Big[ \eps |\nabla_i F|^2
      +(1-\eps)|\nabla_i
      F|^2\bar\chi_{i,T}(\x_i) \Big] \right. \\ \nonumber &&
\!\! +  \sum_{i<j} v_a(|\x_i-\x_j|)|F|^2 - \left. g \sum_{i=1}^N
    |\phi(z_i)|^2 |F|^2\right) \prod_{k=1}^N \phi(z_k)^2
    b_r(\x^\perp_k)^2 d^3\x_k \ . \\ \label{441}
\eeqa 
To bound $Q(F)$ from below, we use again the box method. We divide
$\R$ in $z$-direction into intervals of length $M$, labeled by
$\alpha$, put $N_\alpha$ particles in box $\alpha$, neglect the
interaction between boxes, and minimize over the distribution of the
$N$ particles. This gives a lower bound. More precisely
\beq
\inf_ F Q(F) \geq \inf_{\{N_\alpha\}} \sum_\alpha \inf_{F_\alpha}
Q_\alpha(F_\alpha) \ ,
\eeq
where $F_\alpha=F_\alpha(\x_1,\dots,\x_{N_\alpha})$, and $Q_\alpha$ is
the same as $Q$, but with all the integrations restricted to the box
$\alpha$, and $N$ replaced by $N_\alpha$. The infima are under the
normalization conditions $\int |F|^2 \prod_{k=1}^N \phi(z_k)^2
b_r(\x^\perp_k)^2=1$ and $\int_\alpha |F_\alpha|^2
\prod_{k=1}^{N_\alpha} \phi(z_k)^2 b_r(\x^\perp_k)^2=1$, respectively.

We consider two cases.
Choose $\delta >0$. First, assume that, for all $z\in\alpha$, 
$|\phi(z)|^2\geq \delta N/L$. We use the same method as in
\cite[Eqs.~(5.28)--(5.34)]{LSSY} to get rid of the $\phi^2$ in the
measure $\phi(z)^2 dz$. Let $\pmin$ and $\pmax$ denote the minimal and
maximal value of $\phi$ inside the box $\alpha$, respectively. We
first proceed as in (\ref{329})--(\ref{negl}), and obtain, for
$T\geq r(a/r)^{1/4}=$ radius of $U$, 
\beqa\nonumber 
&&\!\!\!\sum_{i=1}^{N_\alpha} \int_\alpha \left[(1-\eps) |\nabla_i 
F|^2 \bar\chi_{i,T}(\x_i)
  +\half \sum_{j,\, j\neq i} v_a(|\x_i-\x_j|)|F|^2 \right]
\prod_{k=1}^{N_\alpha} \phi(z_k)^2 b_r(\x^\perp_k)^2 d^3 \x_k \\ 
\nonumber &&\!\!\!
\geq \sum_{i=1}^{N_{\alpha}} \int_\alpha \Big[ 
\frac{\pmin^2}{\pmax^2} a'
U(|\x_i-\x_{k(i)}|)\chi_{\B_\delta}(\x^\perp_{k(i)}/r)|F|^2 \Big]
\prod_{k=1}^{N_{\alpha}} \phi(z_k)^2 b_r(\x^\perp_k)^2 d^3 \x_k \ . \\
\label{ensure}
\eeqa 
Here $U$ is given in (\ref{defu}), and $a'$ is given after Eq. 
(\ref{negl}).
Denoting 
\beq 
\widetilde
F(\x_1,\dots,\x_{N_\alpha})=F(\x_1,\dots,\x_{N_\alpha})
\prod_{k=1}^{N_\alpha} \phi(z_k) \ , 
\eeq 
and using 
\beq 
|\nabla_i \widetilde F|^2 \leq 2 |\nabla_i F|^2 \prod_{k=1}^{N_\alpha}
\phi(z_k)^2 + 2 |\widetilde F|^2 \frac {\sup_{z\in \alpha}
  |\phi'|^2}{\pmin^2} \ , 
\eeq 
we get 
\beq \label{off}
 \int_\alpha  |\nabla_i F|^2 \prod_{k=1}^n \phi(z_k)^2
b_r(\x^\perp_k)^2 d^3 \x_k  \geq
 \int_\alpha \left[   \half |\nabla_i \widetilde
F|^2  -\frac{C_{NgL}}{\delta L^2} |\widetilde F|^2
\right] \prod_{k=1}^{N_{\alpha}} b_r(\x^\perp_k)^2 d^3 \x_k \ .  
\eeq 
Here we denoted 
\beq 
C_{NgL}= \frac {L^3}{N} \sup_{z} |\phi'(z)|^2 \ , 
\eeq 
which, by scaling, depends only on $NgL$. 
Estimating
$\phi^2$ by it's maximum in the last term in (\ref{441}), we 
therefore have
\beq
Q_\alpha(F)\geq \widetilde Q_\alpha(\widetilde F) -N_\alpha g\pmax^2 
- \eps
\frac{C_{NgL}}{\delta L^2} N_\alpha \ ,
\eeq
with 
\beq\label{453}
\widetilde Q_\alpha(F) = \sum_{i=1}^{N_\alpha}
\int_\alpha \left( \half \eps |\nabla_i F|^2
     + \frac{\pmin^2}{\pmax^2} a'U(|\x_i-\x_{k(i)}|)|F|^2 \right) 
\prod_{k=1}^N 
  b_r(\x^\perp_k)^2 d^3\x_k  \ .
\eeq
Denote the infimum of $\widetilde Q_\alpha(F)$ over all $F$ (under 
the normalization condition $\int |F|^2
\prod_k b_r(\x^\perp_k)^2 d^3\x_k = 1$) by $\widetilde 
E_\alpha(N_\alpha,
M)$, and choose $\eps=(a/r)^{1/8}$. Looking at the proof of 
Thm.~\ref{finthm}, we see that the lower bound in 
Subsect.~\ref{lobo1} was obtained exactly from an expression like 
(\ref{453}) (compare with (\ref{negl})), except for the additional 
factor $\pmin^2/\pmax^2$. This factor can be estimated by 
\beq 
\frac{\pmin^2}{\pmax^2}\geq 1-
\frac{2 M}{L} \sqrt {\frac {C_{NgL}}\delta } \ . 
\eeq 
Therefore we can apply (\ref{lbthm}), and, in addition, 
Lemma~\ref{L2} from Sunsection 3.4 to estimate $E^{\rm 1D}_{\rm N}$ from below. This gives
\beq\label{446} 
\widetilde E_\alpha (N_\alpha,M) \geq \frac 12
\frac{N_\alpha(N_\alpha-1)}{M} g \left(1- \frac{2 M}{L}\sqrt {\frac 
{C_{NgL}}\delta } \right)
\Big(1-C \nu(N_\alpha,M)\Big) \ ,
\eeq
where 
\beq
\nu(N,M) = N \sqrt{g M} + N \left(\frac ar\right)^{1/8} \left[1 + 
\frac {Nr}{M}\left(\frac ar\right)^{1/8}\right]  \ . 
\eeq
Note that $\nu(N,M)$ is monotone increasing in $N$.

This bound is of no use for large $N_\alpha$, however. Therefore we 
will use
again the box method, as in Subsect.~\ref{subsec43} (see
also~\cite{LY1998}), with small boxes $\ell=M\eta$ for some
$\eta^{-1}\in \N$. The use of Lemma~\ref{L4}, with $L(x)=x[x-1]_+$ 
and $\lambda=4$, implies 
\beq \label{448} 
\widetilde
E_\alpha(N_\alpha,M) \geq \frac 12 \frac{N_\alpha ^2}{M} g 
\left(1-\frac 1{N_\alpha\eta}\right)^2
\left(1-\frac{2M\eta}{L}\sqrt {\frac {C_{NgL}}\delta }\right)\Big(1- 
C\nu(\langle 
4N_\alpha \eta\rangle ,M\eta) \Big) \ .  
\eeq
Let $1\geq \hat\eps \geq 2/N$ such that $\hat\eps N \in \N$, and 
choose 
$\eta=(\hat\eps N)^{-1}$. We estimate $N_\alpha \leq N$ in the last 
term in (\ref{448}), and choose $M=\eps L$ for some $\eps>\hat\eps$.
Minimizing over $N_\alpha$ yields 
\beq\label{naal}
\widetilde E_\alpha(N_\alpha,M) - g \pmin^2 N_\alpha 
\geq -\half g \pmin^4 M {\cal R}  \ ,
\eeq
with 
\beqa\nonumber
{\cal R}&\!\!=&\!\! \left(1+\frac{\hat\eps }{2 \eps 
\delta}\right)^2\left(1-\frac{2\eps}{\hat\eps 
N}\sqrt{\frac{C_{NgL}}\delta}\right)^{-1}\\ 
&&\!\! \cdot \left(1- C \left[ \sqrt\frac{\eps gL}{\hat\eps^3 N} 
+\frac 
1{\hat\eps} \left(\frac ar\right)^{1/8} \left[ 1 + \frac{rN}{\eps L} 
\left(\frac ar\right)^{1/8}\right]\right]\right)^{-1}  \ .
\eeqa

For boxes $\alpha$ where $\pmin^2<N\delta/L$, we just neglect the 
positive
terms in (\ref{441}). This gives 
\beqa\nonumber 
&&\!\!\!\! 
E^{\rm QM}(N,L,r,a) - \left(1-\left(\frac ar\right)^{1/8}\right) 
\sum_{i=1}^{N} \int  |\nabla_i
F|^2\chi_{i,T}(\x_i) \prod_{k=1}^{N}
\phi(z_k)^2 b_r(\x^\perp_k)^2 d^3 \x_k \\ \nonumber &&\!\!\!\!\geq
\frac{Ne^\perp}{r^2} + E^{\rm GP}(N,L,g)+ \frac g2 \int |\phi|^4 
-\left(\frac ar\right)^{1/8}
\frac N{L^2} \frac {C_{NgL}}{\delta} \\ \label{inn}&& \!\!\!\!\quad + 
\sum_{\alpha}\inf_{N_\alpha} \left[ \widetilde E(N_\alpha,M) - g 
\pmax^2 N_\alpha \right] - \delta \frac {N^2 g}{L} \ .  
\eeqa 
Here we neglected the condition $\sum_\alpha N_\alpha =N $ on the 
$N_\alpha$, which can only lower the infimum. 
Moreover, the error in replacing $\pmax^2$ by $\pmin^2$ in the term in
square brackets in (\ref{inn}) is bounded above by
\beq 
2 N g M \sup_{z}
\phi(z)|\phi'(z)| \equiv 2 N^2 g \frac M{L^2} \hat C_{NgL} \ .  
\eeq 
Note that, by scaling,  $\hat C_{NgL}$ depends only on 
$NgL$.

Using (\ref{naal}), 
$\sum_\alpha \pmin^4 M \leq \int \phi^4$ and $\sum_\alpha \pmin^2 M
\leq N$ and dropping the positive second term 
on the left side of (\ref{inn}), we finally obtain
\beqa\nonumber
&&\!\!\!\!\! 
E^{\rm QM}(N,L,r,a)- \frac{Ne^\perp}{r^2}- E^{\rm GP}(N,L,g)
\\ &&\!\!\!\!\!\geq - \frac g2 \int |\phi|^4({\cal R}-1) -
\left(\frac ar\right)^{1/8} \frac 
N{L^2} \frac{C_{NgL}}\delta  - \delta N^2\frac gL  - 2\eps N^2 \frac 
{g}{L} \hat 
C_{NgL} \ . 
\eeqa
Note that $E^{\rm GP}(N,L,g)\geq
\half g \int \phi^4$ and $E^{\rm GP}(N,L,g)\geq c_s N/L^2$ for some
constant $c_s$ depending only on $s$. Hence we see that, for bounded
$NgL$, the parameters $\eps$, $\hat\eps$ and $\delta$ can be chosen
arbitrary small with $a/r$ and $N$ to show the desired lower bound, as
long as $Nr/L\ll (r/a)^{1/4}$ and, in particular, for $Nr/L$
bounded. (Note that $\bar L\sim L$ in Regions 1 and 2.) Here we also
need that $C_{NgL}$ and $\hat C_{NgL}$ are uniformly bounded if
$NgL$ stays bounded, which follows by the same methods as in the proof of
Lemma~\ref{L5} in the Appendix.  For bigger values of $Nr/L$ we have
to proceed differently, using the lower bound (\ref{final}) instead of
(\ref{lbthm}), as we also did in the previous subsection for the lower
bound for Region 3. We omit the details. The results of this
subsection can thus be summarized as
\begin{equation}
\liminf \frac {E^{\rm QM}(N,L,r,a)-Ne^\perp/r^2}{E^{\rm 
GP}(N,L,g)}\geq 1
\end{equation}
in the limit $N\to \infty $ and $r/L\to 0$, uniformly in the
parameters as long as $NgL$ stays bounded.  This finishes the proof of
the lower bound for Regions 1 and 2.

\section{Bose-Einstein Condensation}

In this last section we investigate the question of Bose-Einstein 
condensation  in the ground state. It will be proved to
occur in Regions 1 and 2 but it probably also occurs in part of Region
3; we cannot prove this and it remains an open problem. BEC in the 
ground state means that the one-body density matrix $\gamma(\x, 
\x')$, which is obtained from the ground state wave function $\Psi_0$ 
by
\beq
\gamma(\x,\x')=N\int 
\Psi_0(\x,\x_2,\dots,\x_N)\Psi_0(\x',\x_2,\dots,\x_N)^*d^3\x_2\cdots 
d^3\x_N \ ,
\eeq
factorizes as $N\psi(\x)\psi(\x')$ for some normalized $\psi$ (in the 
$N\to\infty$ limit, of course). This,
in fact, is 100\% condensation. It was proved in \cite{LS} for a 
fixed trap potential in the Gross-Pitaevskii limit, i.e., for both 
$r/L$ and $Na/L$ fixed as $N\to\infty$. Here we extend this result to 
the case $r/L\to 0$ with $NgL$ fixed. The function $\psi $ is the
square-root of the minimizer of the 1D GP functional (\ref{GPfunct})
times the transverse function $b_r(\x^\perp)$.  

BEC is not expected in Regions 4 and 5. Lenard \cite{Lenard} showed
that the largest eigenvalue of $\gamma$ grows only as $N^{1/2}$ for a
homogeneous gas of 1D impenetrable bosons and, according to
\cite{papenbrock} and \cite{forrester}, this holds also for a GT gas
in a harmonic trap. (The exponent 0.59 in \cite{girardeau} can
probably be ascribed to the small number of particles ($N=10$)
considered.)

Our main result about BEC in the ground state is:

\begin{thm}[BEC in Region 2] 
If $N\to \infty$, $r/L\to 0$ with $NaL/r^2$ fixed, then
\beq\label{xx}
\frac{Lr^2}N\gamma(r\x^\perp,Lz;r{\x'}^\perp,Lz')\to 
b(\x^\perp)b({\x'}^\perp)\phi^{\rm GP}(z)\phi^{\rm GP}(z')
\eeq
in trace norm. Here $\phi^{\rm GP}$ is the minimizer of the GP 
functional (\ref{GPfunct}) with $N=1$, $L=1$ and interaction 
parameter $NgL$. 
\end{thm}

\begin{proof}
As in the proof of the energy asymptotics in Region~2, we have to
distinguish the cases $rN/L$ small or large. {F}or simplicity we
consider only the case $rN/L\ll (r/a)^{1/4}$. The case of larger
$rN/L$ can be treated in the same manner, replacing the bound
(\ref{lbthm}) by (\ref{final}) in Subsect.~\ref{ssl12}, and choosing
$T$ in (\ref{defchi}) appropriately.

{F}rom the lower bound to the energy in Subsect.~\ref{ssl12}, together
with the upper bound in Subsect.~\ref{ssu13}, we infer that if
$\Psi_0$ is the ground state of the Hamiltonian $H_{N,L,r,a}$,
$T=r(a/r)^{1/4}$ and $F$ is defined by
\beq
\Psi_0(\x_1,\dots,\x_N)= F(\x_1,\dots,\x_N) \prod_{k=1}^N L^{-1/2} 
\phi^{\rm GP}(z_k/L)  
b_r(\x^\perp_k)  \ ,
\eeq
then 
\beq
\lim_{N\to\infty} \frac{L^2}{N} 
\sum_{i=1}^{N} \int |\nabla_i F|^2\chi_{i,T}(\x_i) \prod_{k=1}^{N}
L^{-1} \phi^{\rm GP}(z_k/L)^2 b_r(\x^\perp_k)^2 d^3 \x_k = 0 
\eeq
in the limit $N\to\infty$, $r/L\to 0$ with $NgL$ fixed. Here
$\chi_{i,T}$ is given in (\ref{defchi}). Note that in this limit
$NT^3/(r^2 L)=Nr/L (a/r)^{3/4}\to 0$, i.e., the volume of the set
where $\chi_{i,T}$ is zero is small compared to the total volume $r^2
L$. Eq. (\ref{xx}) now follows, using the methods of
\cite{LS}.
\end{proof}

\appendix
\section{Appendix: Proof of auxiliary Lemmas}

\begin{proof}[Proof of Lemma~\ref{L5}]
Without restriction we may assume that $V^\perp\geq 0$. 
The existence, uniqueness and positivity of a minimizer $\phi_p$ are 
standard (cf., e.g., \cite{lsy1}). From the variational equation
\beq
\left(-\Delta^\perp + V^\perp(\x_\perp) + 2p 
|\phi_p(\x^\perp)|^2\right) \phi_p(\x^\perp) = \mu_p \phi_p(\x^\perp)
\eeq
we infer that, for $K(\x^\perp-\y^\perp)$ the integral kernel of 
$(-\Delta^\perp+1)^{-1}$, 
\beq
\phi_p(\x^\perp)=\int K(\x^\perp-\y^\perp) 
\left(\mu_p+1-V^\perp(\y^\perp)-2p|\phi_p(\y^\perp)|^2\right) 
\phi_p(\y^\perp) d^2\y^\perp \ .
\eeq
Using positivity of $V^\perp$, $\phi_p$ and $K$, as well as the 
normalization of $\phi_p$,  we obtain the bound
\beqa\nonumber
\phi_p(\x^\perp)e^{|\x^\perp|} &\leq& \sup_{\x^\perp}  
\int_{V^\perp(\y^\perp)\leq \mu_p+1} e^{|\x^\perp|} 
K(\x^\perp-\y^\perp) \left(\mu_p+1\right) \phi_p(\y^\perp)d^2\y^\perp 
\\ \nonumber&\leq& (\mu_p+1) \sup_{\x^\perp}\left( 
\int_{V^\perp(\y^\perp)\leq \mu_p+1} \left| e^{|\x^\perp|} 
K(\x^\perp-\y^\perp)\right|^2 d^2\y^\perp \right)^{1/2} \equiv C_p \ 
.\\ \label{bound1}
\eeqa
Since $\mu_p$ is uniformly bounded for $p$ in a bounded interval,
so is $C_p$. 
Moreover,
\beqa\nonumber
|\nabla^\perp \phi_p(\x^\perp)| &\leq& \int \left|\nabla^\perp 
K(\x^\perp-\y^\perp)\right| \left| \mu_p +1 -V^\perp(\y^\perp)-2p 
|\phi_p(\y^\perp)|^2\right| \phi_p(\y^\perp) d^2\y^\perp \\&\leq& 
\widetilde C_p \|\nabla^\perp K\|_1\ ,
\eeqa
with
\beq
\widetilde C_p = \sup_{\y^\perp} \left| \mu_p +1 
-V^\perp(\y^\perp)-2p |\phi_p(\y^\perp)|^2\right| \phi_p(\y^\perp) \ ,
\eeq
which is finite and uniformly bounded for bounded $p$ because of 
(\ref{bound1}) and the fact that that $V^\perp$ is 
polynomially bounded at infinity by assumption. 
\end{proof}

\begin{proof}[Proof of Lemma~\ref{L2}]
We write (\ref{13}) as
\beq\label{133}
H_{n,g}=\sum_{j=1}^n-\half \partial_j^2 + \half \sum_{i\neq j}\left[  
-\frac 1{n-1} \partial_j^2 + g \delta(z_i-z_j) \right] \ .
\eeq
To bound this expression from below, we want to use Temple's 
inequality, and for this purpose we have to smear out the 
$\delta$-function interaction. We use the following Lemma (compare 
with \cite[Lemma~6.3]{BSY}), whose proof can be found below.

\begin{lem}\label{Lh}
Let $\partial^2_z$ denote the Neumann Laplacian on an interval
$[0,\ell]$, let $z_0\in (0,\ell)$, and
choose positive numbers $A$, $B$ and $\alpha$ such that $R\equiv B
\arctan(BA/2\alpha)\leq \min\{z_0,\ell-z_0\}$. Then
\beq\label{a2}
-\alpha \partial_z^2 + A \delta(z-z_0) - \frac 
\alpha{B^2}\theta(R-|z-z_0|)\geq 0 \ .
\eeq
\end{lem}

We apply this result to the operator in square brackets in 
(\ref{133}). This gives
\beq
H_{n,g}\geq \sum_{j=1}^n-\half \partial_j^2 + \half \sum_{i\neq j}
\frac 1{(n-1)B^2} \theta(R-|z_j-z_i|) \chi_{[R,\ell-R])}(z_i) \ ,
\eeq
with $R=B \arctan(Bg(n-1)/2)$ and $B>0$ arbitrary. Temple's 
inequality (\ref{temple}) implies that, for
\beq
U(z_1,\dots,z_n)= \half \sum_{i\neq j}
\frac 1{(n-1)B^2} \theta(R-|z_j-z_i|) \chi_{[R,\ell-R])}(z_i) 
\eeq
and $\langle 
U^k\rangle=\ell^{-n} \int U^k dz_1\dots dz_n$, 
\beq
E_{\rm N}^{\rm 1D}(n,\ell,g)\geq \langle U\rangle \left( 
1-\frac{\langle U^2\rangle}{\langle U\rangle} \frac 1{\half 
\pi^2/\ell^2 - \langle U\rangle}\right) \ ,
\eeq
provided the term in the denominator is positive. By Schwarz'
inequality,
\beq
\langle U^2 \rangle \leq \frac{n}{2B^2} \langle U \rangle \ .
\eeq
Moreover,
\beq
\langle U \rangle=\frac 12 \frac n{B^2} \frac 1{\ell^2}\int_0^\ell dz 
\int_R^{\ell-R} dw \theta (R-|z-w|) = \frac n{B^2}\frac R{\ell^2} 
(\ell - 2R) \ .
\eeq
Using $x\geq \arctan(x)\geq x(1-x/3)$ for $x\geq 0$, this leads to 
the 
estimate
\beq
E^{\rm 1D}_{\rm N}(n,\ell,g)\geq \half n(n-1) \frac g\ell 
\left(1-\frac{B^2 g n}\ell - \frac{Bgn}{6} 
\right)\left(1-\frac{n}{B^2(\pi^2/\ell^2-n(n-1)g/\ell)}\right) \ ,
\eeq
under the assumption $\pi^2/\ell^2> n(n-1)g/\ell$. Choosing 
$B=\ell(\ell g\pi^2)^{-1/4}$, this gives, for ${\cal R}= (n^2 \ell 
g)^{1/2}/\pi$,
\beq
E^{\rm 1D}_{\rm N}(n,\ell,g)\geq \half n(n-1) \frac g\ell 
\left(1-{\cal
R}-\frac \pi{6n^{1/2}} {\cal R}^{3/2}\right)\left(1-\frac {\cal
R}{1-{\cal R}^2}\right)
\eeq
and proves the desired lower bound.
\end{proof}

\begin{proof}[Proof of Lemma~\ref{Lh}]
Let $h$ denote the operator on the left side of (\ref{a2}). The 
function
\beq
f(z)= \left\{ \begin{array} {ll}  \cos\Big((R-|z-z_0|)/B\Big) & {\rm 
for\ } 0\leq |z-z_0|\leq R\\ 1 & {\rm otherwise } \end{array} \right.
\eeq
fulfills the Schr\"odinger equation $hf=0$, and since it is 
positive, it must be the ground state of $h$. Hence $h\geq 0$. 
\end{proof}

\begin{proof}[Proof of Lemma~\ref{L4}]
Let $p=\langle \lambda N/M\rangle$. For $n\leq p$ we use the estimate
\beq
E(n)\geq L(n)K(p) \ ,
\eeq
which follows from monotonicity of $K$. For $n\geq p$, we use 
superadditivity, which implies that
\beq
E(n)\geq [n/p] E(p) \geq \frac n{2p} L(p)K(p) \ ,
\eeq
where $[x]$ denotes the integer part of $x$. Now let $t=\sum_{n<p} 
c_n n$. By convexity of $L$ and the fact that $L(0)=0$, 
\beq
\sum_{n<p} c_n L(n) \geq M L(t/M) \ .
\eeq
Hence
\beq\label{a10}
\sum_{n\geq 0} c_n E(n) \geq K(p) \left[ M L(t/M) + (N-t) \frac 
{L(p)}{2p} \right] \ .
\eeq
To obtain a lower bound, we have to minimize the right side over all 
$0\leq t \leq N$. Note that (\ref{a10}) is convex in $t$, hence the 
minimum is either taken at $t=N$, or, if it is taken at some $t_0<N$, 
it's right derivative at $t_0$ has to be positive. Using (\ref{lpr}) 
this leads to 
\beq
\frac{L(\lambda t_0/M)}{2\lambda t_0/M}\geq \frac{L(p)}{2p} \ .
\eeq
By our assumptions on $L$, $L(x)/x$ is monotone increasing, and hence 
$t_0\geq pM/\lambda\geq N$. Thus we can set $t=N$ in (\ref{a10}), and 
obtain the desired lower bound.
\end{proof}

\end{document}